\documentclass{article}

\usepackage{mathtools}
\usepackage{booktabs} 

\usepackage{comment}
\usepackage{tikz}
\usetikzlibrary{matrix,arrows,decorations.pathmorphing,patterns,positioning}
\usepackage{amssymb}
\usepackage{ifthen}
\usepackage{color}
\usepackage{colortbl}
\usepackage{rotating}
\usepackage{multirow}
\usepackage{multicol}
\usepackage{url}
\usepackage{arydshln}

\usepackage{hhline}

\usepackage{pgfplots}

\usepackage{mathtools}

\usetikzlibrary{external}
\usepackage{wrapfig}
\usepackage{subcaption}

\usepackage{caption}

\usepackage{array}
\usepackage{framed}

\usepackage{balance}

\usepackage{epsfig}
\usepackage{amssymb}

\usepackage{algorithmic}

\usepackage{algorithm}

\usepackage[normalem]{ulem}

\usepackage{ifthen}
\usepackage{boxedminipage}
\usepackage{fancyvrb}

\usepackage{soul}
\usepackage{graphicx}
\usepackage{graphics}
\usepackage{moreverb}
\usepackage{enumerate}

\usepackage{enumitem}

\usepackage{tabularx,booktabs}

\usetikzlibrary{arrows}

\usepackage{xspace}

\usepackage{stmaryrd}

\usepackage[absolute]{textpos}
\TPMargin{5pt}

\usepackage{listings}
\usepackage{xcolor}
\usepackage{fancyvrb}
\fvset{tabsize=2}
\definecolor{mygreen}{rgb}{0,0.6,0}
\usepackage{afterpage}

\usepackage[mathscr]{eucal}

\usepackage{pdflscape}

\usepackage[export]{adjustbox}
\usepackage{appendix}

\definecolor{colorbrewer1}{RGB}{228,26,28}
\definecolor{colorbrewer2}{RGB}{55,126,184}
\definecolor{colorbrewer3}{RGB}{77,175,74}
\definecolor{colorbrewer4}{RGB}{152,78,163}
\definecolor{colorbrewer5}{RGB}{255,127,0}
\definecolor{colorbrewer7}{RGB}{166,86,40}
\definecolor{colorbrewer8}{RGB}{247,129,191}
\definecolor{colorbrewer9}{RGB}{153,153,153}

\definecolor{darkgreen}{rgb}{0,0.5,0}

\definecolor{darkred}{rgb}{0.44,0,0}
\definecolor{darkgreen}{rgb}{0,0.44,0}
\definecolor{darkblue}{rgb}{0,0,0.44}
\definecolor{enrique}{rgb}{0,0,0}

\pgfplotscreateplotcyclelist{jianyu2color}{
	colorbrewer1, mark=*,\\
	colorbrewer2, mark=square*,\\
	colorbrewer3, mark=triangle*,\\
	colorbrewer4, mark=x,\\
	colorbrewer5, mark=*,\\
	colorbrewer7, mark=triangle*,\\
	colorbrewer8, mark=+,\\
}

\pgfplotscreateplotcyclelist{jianyu0color}{%
ultra thick,colorbrewer1,every mark/.append style={fill=colorbrewer1},mark=triangle*\\%
orange,every mark/.append style={fill=orange},mark=square*\\%
cyan,every mark/.append style={fill=cyan},mark=otimes*\\%
red!70!white,mark=star\\%
lime!80!black,every mark/.append style={fill=lime},mark=diamond*\\%
red,densely dashed,every mark/.append style={solid,fill=red!80!black},mark=*\\%
thick,yellow!60!black,every mark/.append style={solid,fill=yellow!80!black},mark=square*\\%
purple,every mark/.append style={solid,fill=gray},mark=otimes*\\%
blue,densely dashed,mark=star,every mark/.append style=solid\\%
red,densely dashed,every mark/.append style={solid,fill=red!80!black},mark=diamond*\\%
thick,colorbrewer4,mark=x\\%
blue,every mark/.append style={fill=blue!80!black},mark=*\\%
red,every mark/.append style={fill=red!80!black},mark=square*\\%
brown!60!black,every mark/.append style={fill=brown!80!black},mark=diamond*\\%
thick,darkgreen,mark=star\\%
blue,every mark/.append style={fill=blue!80!black},mark=diamond*\\%
red,densely dashed,every mark/.append style={solid,fill=red!80!black},mark=*\\%
brown!60!black,densely dashed,every mark/.append style={
solid,fill=brown!80!black},mark=square*\\%
darkred,densely dashed,every mark/.append style={solid,fill=gray},mark=otimes*\\%
blue,densely dashed,mark=oplus,every mark/.append style=solid\\%
colorbrewer5,densely dashed,every mark/.append style={solid,fill=colorbrewer5},mark=diamond*\\%
darkgreen,densely dashed,mark=+,every mark/.append style=solid\\%
thick,colorbrewer7,mark=triangle*\\%
thick,black\\%
}

\pgfplotscreateplotcyclelist{jianyu0colorscatter}{%
thick,colorbrewer1,every mark/.append style={fill=colorbrewer1},mark=triangle*,only marks\\%
thick,orange,every mark/.append style={fill=orange},mark=square*,only marks\\%
thick, lime!80!black,every mark/.append style={fill=lime!80!black},mark=pentagon*,only marks\\%
thick,blue,every mark/.append style={fill=blue!80!black},mark=triangle*,only marks\\%
thick,cyan,every mark/.append style={fill=cyan},mark=square*,only marks\\%
thick,darkgreen,mark=pentagon*,only marks\\%
thick,yellow!60!black,every mark/.append style={solid,fill=yellow!80!black},mark=square*, only marks\\%
thick,purple,every mark/.append style={solid,fill=gray},mark=otimes*, only marks\\%
thick,blue,densely dashed,mark=star,every mark/.append style=solid, only marks\\%
thick,red,densely dashed,every mark/.append style={solid,fill=red!80!black},mark=diamond*, only marks\\%
thick,colorbrewer4,mark=x\\%
thick,blue,every mark/.append style={fill=blue!80!black},mark=*\\%
}

\pgfplotscreateplotcyclelist{jianyucolor}{%
ultra thick,colorbrewer1,every mark/.append style={fill=colorbrewer1},mark=triangle*\\%
orange,every mark/.append style={fill=orange},mark=square*\\%
cyan,every mark/.append style={fill=cyan},mark=otimes*\\%
red!70!white,mark=star\\%
lime!80!black,every mark/.append style={fill=lime},mark=diamond*\\%
red,densely dashed,every mark/.append style={solid,fill=red!80!black},mark=*\\%
thick,yellow!60!black,every mark/.append style={solid,fill=yellow!80!black},mark=square*\\%
purple,every mark/.append style={solid,fill=gray},mark=otimes*\\%
blue,densely dashed,mark=star,every mark/.append style=solid\\%
red,densely dashed,every mark/.append style={solid,fill=red!80!black},mark=diamond*\\%
thick,colorbrewer4,mark=x\\%
blue,every mark/.append style={fill=blue!80!black},mark=*\\%
red,every mark/.append style={fill=red!80!black},mark=square*\\%
brown!60!black,every mark/.append style={fill=brown!80!black},mark=diamond*\\%
thick,darkgreen,mark=star\\%
blue,every mark/.append style={fill=blue!80!black},mark=diamond*\\%
red,densely dashed,every mark/.append style={solid,fill=red!80!black},mark=*\\%
brown!60!black,densely dashed,every mark/.append style={
solid,fill=brown!80!black},mark=square*\\%
darkred,densely dashed,every mark/.append style={solid,fill=gray},mark=otimes*\\%
blue,densely dashed,mark=oplus,every mark/.append style=solid\\%
colorbrewer5,densely dashed,every mark/.append style={solid,fill=colorbrewer5},mark=diamond*\\%
darkgreen,densely dashed,mark=+,every mark/.append style=solid\\%
thick,colorbrewer7,mark=triangle*\\%
thick,black,densely dashed\\%
thick,black\\%
}

\pgfplotscreateplotcyclelist{jianyucolorscatter}{%
thin,black,mark=o,only marks\\%
thick,colorbrewer1,every mark/.append style={fill=colorbrewer1},mark=triangle*,only marks\\%
thick,orange,every mark/.append style={fill=orange},mark=square*,only marks\\%
thick, lime!80!black,every mark/.append style={fill=lime!80!black},mark=pentagon*,only marks\\%
thick,blue,every mark/.append style={fill=blue!80!black},mark=triangle*,only marks\\%
thick,cyan,every mark/.append style={fill=cyan},mark=square*,only marks\\%
thick,darkgreen,mark=pentagon*,only marks\\%
thick,yellow!60!black,every mark/.append style={solid,fill=yellow!80!black},mark=square*, only marks\\%
thick,purple,every mark/.append style={solid,fill=gray},mark=otimes*, only marks\\%
thick,blue,densely dashed,mark=star,every mark/.append style=solid, only marks\\%
thick,red,densely dashed,every mark/.append style={solid,fill=red!80!black},mark=diamond*, only marks\\%
thick,colorbrewer4,mark=x\\%
thick,blue,every mark/.append style={fill=blue!80!black},mark=*\\%
red,every mark/.append style={fill=red!80!black},mark=square*\\%
brown!60!black,every mark/.append style={fill=brown!80!black},mark=diamond*\\%
blue,every mark/.append style={fill=blue!80!black},mark=diamond*\\%
thick,red,densely dashed,every mark/.append style={solid,fill=red!80!black},mark=*, only marks\\%
red,densely dashed,every mark/.append style={solid,fill=red!80!black},mark=*\\%
brown!60!black,densely dashed,every mark/.append style={
solid,fill=brown!80!black},mark=square*\\%
darkred,densely dashed,every mark/.append style={solid,fill=gray},mark=otimes*\\%
blue,densely dashed,mark=oplus,every mark/.append style=solid\\%
colorbrewer5,densely dashed,every mark/.append style={solid,fill=colorbrewer5},mark=diamond*\\%
darkgreen,densely dashed,mark=+,every mark/.append style=solid\\%
thick,colorbrewer7,mark=triangle*\\%
thick,black,densely dashed\\%
thick,black\\%
}

\pgfplotscreateplotcyclelist{jianyucolordot}{%
black,mark=-,only marks\\%
colorbrewer1,mark=o,only marks\\%
brown!60!black,mark=+,only marks\\%
darkred,mark=x,only marks\\%
colorbrewer4,mark=pentagon,only marks\\%
blue,mark=triangle,only marks\\%
purple,mark=square,only marks\\%
darkgreen,mark=diamond,only marks\\%
thick,cyan,every mark/.append style={solid,fill=gray}\\%
thick,blue,densely dashed,mark=star,every mark/.append style=solid\\%
thick,red,densely dashed,every mark/.append style={solid,fill=red!80!black},mark=diamond*, only marks\\%
thick,yellow!60!black,mark=x\\%
thick,blue,every mark/.append style={fill=blue!80!black},mark=*\\%
red,every mark/.append style={fill=red!80!black},mark=square*\\%
brown!60!black,every mark/.append style={fill=brown!80!black},mark=diamond*\\%
blue,every mark/.append style={fill=blue!80!black},mark=diamond*\\%
thick,red,densely dashed,every mark/.append style={solid,fill=red!80!black},mark=*, only marks\\%
red,densely dashed,every mark/.append style={solid,fill=red!80!black},mark=*\\%
orange,densely dashed,every mark/.append style={
solid,fill=brown!80!black},mark=square*\\%
lime!80!black,densely dashed,every mark/.append style={solid,fill=gray},mark=otimes*\\%
blue,densely dashed,mark=oplus,every mark/.append style=solid\\%
colorbrewer5,densely dashed,every mark/.append style={solid,fill=colorbrewer5},mark=diamond*\\%
darkgreen,densely dashed,mark=+,every mark/.append style=solid\\%
thick,colorbrewer7,mark=triangle*\\%
thick,black,densely dashed\\%
thick,black\\%
}

\pgfplotscreateplotcyclelist{jianyucolorbar}{%
fill=black\\%
fill=colorbrewer1\\%
fill=orange\\%
fill=lime!80!black\\%
fill=blue\\%
fill=cyan\\%
fill=darkgreen\\%
fill=yellow!60!black\\%
thick,purple,every mark/.append style={solid,fill=gray}\\%
thick,blue,densely dashed,mark=star,every mark/.append style=solid\\%
thick,red,densely dashed,every mark/.append style={solid,fill=red!80!black},mark=diamond*, only marks\\%
thick,colorbrewer4,mark=x\\%
thick,blue,every mark/.append style={fill=blue!80!black},mark=*\\%
red,every mark/.append style={fill=red!80!black},mark=square*\\%
brown!60!black,every mark/.append style={fill=brown!80!black},mark=diamond*\\%
blue,every mark/.append style={fill=blue!80!black},mark=diamond*\\%
thick,red,densely dashed,every mark/.append style={solid,fill=red!80!black},mark=*, only marks\\%
red,densely dashed,every mark/.append style={solid,fill=red!80!black},mark=*\\%
brown!60!black,densely dashed,every mark/.append style={
solid,fill=brown!80!black},mark=square*\\%
darkred,densely dashed,every mark/.append style={solid,fill=gray},mark=otimes*\\%
blue,densely dashed,mark=oplus,every mark/.append style=solid\\%
colorbrewer5,densely dashed,every mark/.append style={solid,fill=colorbrewer5},mark=diamond*\\%
darkgreen,densely dashed,mark=+,every mark/.append style=solid\\%
thick,colorbrewer7,mark=triangle*\\%
thick,black,densely dashed\\%
thick,black\\%
}

\pgfplotscreateplotcyclelist{jianyucolorlineratio}{%
thick,black,mark=*\\%
thick,colorbrewer1,every mark/.append style={fill=colorbrewer1},mark=triangle*\\%
thick,orange,every mark/.append style={fill=orange},mark=square*\\%
thick, lime!80!black,every mark/.append style={fill=lime!80!black},mark=pentagon*\\%
thick,blue,every mark/.append style={fill=blue!80!black},mark=triangle*\\%
thick,cyan,every mark/.append style={fill=cyan},mark=square*\\%
thick,darkgreen,mark=pentagon*\\%
thick,yellow!60!black,every mark/.append style={solid,fill=yellow!80!black}\\%
thick,purple,every mark/.append style={solid,fill=gray}\\%
thick,blue,densely dashed,mark=star,every mark/.append style=solid\\%
thick,red,densely dashed,every mark/.append style={solid,fill=red!80!black},mark=diamond*, only marks\\%
thick,colorbrewer4,mark=x\\%
thick,blue,every mark/.append style={fill=blue!80!black},mark=*\\%
red,every mark/.append style={fill=red!80!black},mark=square*\\%
brown!60!black,every mark/.append style={fill=brown!80!black},mark=diamond*\\%
blue,every mark/.append style={fill=blue!80!black},mark=diamond*\\%
thick,red,densely dashed,every mark/.append style={solid,fill=red!80!black},mark=*, only marks\\%
red,densely dashed,every mark/.append style={solid,fill=red!80!black},mark=*\\%
brown!60!black,densely dashed,every mark/.append style={
solid,fill=brown!80!black},mark=square*\\%
darkred,densely dashed,every mark/.append style={solid,fill=gray},mark=otimes*\\%
blue,densely dashed,mark=oplus,every mark/.append style=solid\\%
colorbrewer5,densely dashed,every mark/.append style={solid,fill=colorbrewer5},mark=diamond*\\%
darkgreen,densely dashed,mark=+,every mark/.append style=solid\\%
thick,colorbrewer7,mark=triangle*\\%
thick,black,densely dashed\\%
thick,black\\%
}

\pgfplotscreateplotcyclelist{jianyucolorline}{%
thin,black\\%
thick,colorbrewer1,every mark/.append style={fill=colorbrewer1}\\%
thick,orange,every mark/.append style={fill=orange}\\%
thick, lime!80!black,every mark/.append style={fill=lime!80!black}\\%
thick,blue,every mark/.append style={fill=blue!80!black}\\%
thick,cyan,every mark/.append style={fill=cyan}\\%
thick,darkgreen\\%
thick,yellow!60!black,every mark/.append style={solid,fill=yellow!80!black}\\%
thick,purple,every mark/.append style={solid,fill=gray}\\%
thick,blue,densely dashed,mark=star,every mark/.append style=solid\\%
thick,red,densely dashed,every mark/.append style={solid,fill=red!80!black},mark=diamond*, only marks\\%
thick,colorbrewer4,mark=x\\%
thick,blue,every mark/.append style={fill=blue!80!black},mark=*\\%
red,every mark/.append style={fill=red!80!black},mark=square*\\%
brown!60!black,every mark/.append style={fill=brown!80!black},mark=diamond*\\%
blue,every mark/.append style={fill=blue!80!black},mark=diamond*\\%
thick,red,densely dashed,every mark/.append style={solid,fill=red!80!black},mark=*, only marks\\%
red,densely dashed,every mark/.append style={solid,fill=red!80!black},mark=*\\%
brown!60!black,densely dashed,every mark/.append style={
solid,fill=brown!80!black},mark=square*\\%
darkred,densely dashed,every mark/.append style={solid,fill=gray},mark=otimes*\\%
blue,densely dashed,mark=oplus,every mark/.append style=solid\\%
colorbrewer5,densely dashed,every mark/.append style={solid,fill=colorbrewer5},mark=diamond*\\%
darkgreen,densely dashed,mark=+,every mark/.append style=solid\\%
thick,colorbrewer7,mark=triangle*\\%
thick,black,densely dashed\\%
thick,black\\%
}

\pgfplotscreateplotcyclelist{jianyucolordistline}{%
thin,black,mark=*\\%
thin,colorbrewer1,every mark/.append style={fill=colorbrewer1},mark=triangle*\\%
thin,orange,every mark/.append style={fill=orange},mark=square*\\%
thin,lime!80!black,every mark/.append style={fill=lime!80!black},mark=pentagon*\\%
thin,blue,every mark/.append style={fill=blue!80!black},mark=triangle*\\%
thin,cyan,every mark/.append style={fill=cyan},mark=square*\\%
thin,darkgreen,mark=pentagon*\\%
densely dashed,black,mark=x\\%
thick,yellow!60!black,every mark/.append style={solid,fill=yellow!80!black}\\%
thick,purple,every mark/.append style={solid,fill=gray}\\%
thick,blue,densely dashed,mark=star,every mark/.append style=solid\\%
thick,red,densely dashed,every mark/.append style={solid,fill=red!80!black},mark=diamond*, only marks\\%
thick,colorbrewer4,mark=x\\%
thick,blue,every mark/.append style={fill=blue!80!black},mark=*\\%
red,every mark/.append style={fill=red!80!black},mark=square*\\%
brown!60!black,every mark/.append style={fill=brown!80!black},mark=diamond*\\%
blue,every mark/.append style={fill=blue!80!black},mark=diamond*\\%
thick,red,densely dashed,every mark/.append style={solid,fill=red!80!black},mark=*, only marks\\%
red,densely dashed,every mark/.append style={solid,fill=red!80!black},mark=*\\%
brown!60!black,densely dashed,every mark/.append style={
solid,fill=brown!80!black},mark=square*\\%
darkred,densely dashed,every mark/.append style={solid,fill=gray},mark=otimes*\\%
blue,densely dashed,mark=oplus,every mark/.append style=solid\\%
colorbrewer5,densely dashed,every mark/.append style={solid,fill=colorbrewer5},mark=diamond*\\%
darkgreen,densely dashed,mark=+,every mark/.append style=solid\\%
thick,colorbrewer7,mark=triangle*\\%
thick,black,densely dashed\\%
thick,black\\%
}

\pgfplotscreateplotcyclelist{jianyucolorboth}{%
thin,black,mark=o,only marks\\%
thick,colorbrewer1,every mark/.append style={fill=colorbrewer1},mark=triangle*,only marks\\%
thick,orange,every mark/.append style={fill=orange},mark=square*,only marks\\%
thick, lime!80!black,every mark/.append style={fill=lime!80!black},mark=pentagon*,only marks\\%
thick,blue,every mark/.append style={fill=blue!80!black},mark=triangle*,only marks\\%
thick,cyan,every mark/.append style={fill=cyan},mark=square*,only marks\\%
thick,darkgreen,mark=pentagon*,only marks\\
thin,black\\%
thick,colorbrewer1,every mark/.append style={fill=colorbrewer1}\\%
thick,orange,every mark/.append style={fill=orange}\\%
thick, lime!80!black,every mark/.append style={fill=lime!80!black}\\%
thick,blue,every mark/.append style={fill=blue!80!black}\\%
thick,cyan,every mark/.append style={fill=cyan}\\%
thick,darkgreen\\
thick,yellow!60!black,every mark/.append style={solid,fill=yellow!80!black}\\%
thick,purple,every mark/.append style={solid,fill=gray}\\%
thick,blue,densely dashed,mark=star,every mark/.append style=solid\\%
thick,red,densely dashed,every mark/.append style={solid,fill=red!80!black},mark=diamond*, only marks\\%
thick,colorbrewer4,mark=x\\%
thick,blue,every mark/.append style={fill=blue!80!black},mark=*\\%
red,every mark/.append style={fill=red!80!black},mark=square*\\%
brown!60!black,every mark/.append style={fill=brown!80!black},mark=diamond*\\%
blue,every mark/.append style={fill=blue!80!black},mark=diamond*\\%
thick,red,densely dashed,every mark/.append style={solid,fill=red!80!black},mark=*, only marks\\%
red,densely dashed,every mark/.append style={solid,fill=red!80!black},mark=*\\%
brown!60!black,densely dashed,every mark/.append style={
solid,fill=brown!80!black},mark=square*\\%
darkred,densely dashed,every mark/.append style={solid,fill=gray},mark=otimes*\\%
blue,densely dashed,mark=oplus,every mark/.append style=solid\\%
colorbrewer5,densely dashed,every mark/.append style={solid,fill=colorbrewer5},mark=diamond*\\%
darkgreen,densely dashed,mark=+,every mark/.append style=solid\\%
thick,colorbrewer7,mark=triangle*\\%
thick,black,densely dashed\\%
thick,black\\%
}

\pgfplotscreateplotcyclelist{jianyu3color}{%
black,densely dashed\\%
black\\%
colorbrewer1,every mark/.append style={fill=colorbrewer1},mark=triangle*\\%
orange,every mark/.append style={fill=orange},mark=square*\\%
cyan,every mark/.append style={fill=cyan},mark=otimes*\\%
red!70!white,mark=star\\%
brown!60!black,every mark/.append style={fill=brown!80!black},mark=otimes*\\%
purple,every mark/.append style={solid,fill=gray},mark=otimes*\\%
very thick, blue,every mark/.append style={fill=lime},mark=diamond*\\%
very thick, darkgreen,mark=+,every mark/.append style=solid\\%
}

\pgfplotscreateplotcyclelist{jianyu4color}{%
black,densely dashed,mark=triangle*\\%
black,mark=triangle*\\%
colorbrewer1,densely dashed,mark=triangle*\\%
thick,colorbrewer1,mark=triangle*\\%
black,densely dashed,mark=square*\\%
black,mark=square*\\%
blue,densely dashed,mark=square*\\%
thick,blue,mark=square*\\%
black,densely dashed,mark=*\\%
black,mark=*\\%
darkgreen,densely dashed,mark=*\\%
thick,darkgreen,mark=*\\%
orange,mark=square\\%
cyan,mark=otimes\\%
purple,mark=star\\%
}

\pgfplotscreateplotcyclelist{jianyu5color}{%
black,densely dashed\\%
black\\%
darkgreen,densely dashed,mark=*\\%
thick,darkgreen,mark=*\\%
blue,densely dashed,mark=square*\\%
thick,blue,mark=square*\\%
colorbrewer1,densely dashed,mark=triangle*\\%
thick,colorbrewer1,mark=triangle*\\%
}

\pgfplotscreateplotcyclelist{jianyu6color}{%
black,densely dashed\\%
black\\%
very thick,densely dashed,colorbrewer1,every mark/.append style={fill=colorbrewer1},mark=triangle*\\%
very thick,orange,every mark/.append style={fill=orange},mark=square*\\%
cyan,every mark/.append style={fill=cyan},mark=otimes*\\%
red!70!white,mark=star\\%
brown!60!black,every mark/.append style={fill=brown!80!black},mark=otimes*\\%
purple,every mark/.append style={solid,fill=gray},mark=otimes*\\%
blue,every mark/.append style={fill=lime},mark=diamond*\\%
darkgreen,mark=+,every mark/.append style=solid\\%
}

\pgfplotscreateplotcyclelist{jianyugpucolor}{%
black\\%
black,densely dashed\\%
thick,darkgreen,mark=*\\%
thick,darkgreen,densely dashed,mark=*\\%
thick,colorbrewer1,mark=triangle*\\%
thick,blue,densely dashed,mark=square*\\%
thick,blue,mark=square*\\%
thick,colorbrewer1,densely dashed,mark=triangle*\\%
}

\definecolor{excel1}{RGB}{65,110,166}
\definecolor{excel2}{RGB}{185,86,80}
\definecolor{excel3}{RGB}{151,177,92}
\definecolor{excel4}{RGB}{129,106,158}
\definecolor{excel5}{RGB}{74,166,188}
\definecolor{excel6}{RGB}{227,148,70}

\pgfplotscreateplotcyclelist{jianyucutlasscolor}{%
thick,excel1,mark=diamond*\\%
thick,excel2,mark=square*\\%
thick,excel3,mark=triangle*\\%
thick,excel4,mark=x\\%
thick,excel5,mark=pentagon*\\%
thick,excel6,mark=*\\%
black,densely dashed\\%
}

\pgfplotscreateplotcyclelist{jianyumodelcolor}{%
black\\%
black,densely dashed\\%
thick,darkgreen,mark=*\\%
thick,orange,mark=*\\%
thick,colorbrewer1,mark=triangle*\\%
thick,blue,mark=square*\\%
}

\pgfplotscreateplotcyclelist{jianyustability222color}{%
densely dashed,colorbrewer1,mark=triangle*\\%
densely dashed,blue,mark=square*\\%
densely dashed,darkgreen,mark=*\\%
thick,colorbrewer1,mark=triangle*\\%
thick,blue,mark=square*\\%
thick,darkgreen,mark=*\\%
densely dashdotdotted,colorbrewer1,mark=triangle*\\%
densely dashdotdotted,blue,mark=square*\\%
densely dashdotdotted,darkgreen,mark=*\\%
}

\pgfplotscreateplotcyclelist{jianyuhugecolor}{%
thick,black,mark=*\\%
thick,blue,mark=triangle*\\%
thick,colorbrewer1,mark=square*\\%
thick,orange,every mark/.append style={fill=orange},mark=pentagon*\\%
thick,darkgreen,mark=diamond\\%
thick,cyan,every mark/.append style={fill=cyan},mark=square*\\%
thick,black,densely dashed,mark=*\\%
thick,blue,densely dashed,mark=square*\\%
thick,colorbrewer1,densely dashed,mark=triangle*\\%
thick,orange,every mark/.append style={fill=orange},densely dashed,mark=pentagon*\\%
thick,darkgreen,densely dashed,mark=diamond\\%
thick,cyan,every mark/.append style={fill=cyan},densely dashed,mark=square*\\%
}

\def\gap{\hspace*{.15in}}

\newcommand{\Strassen}{Strassen}

\newcommand{\figref}[1]{Figure~\ref{#1}}

\newcommand{\tabref}[1]{Figure~\ref{#1}}

\newcommand{\secref}[1]{Section~\ref{#1}}
\newcommand{\secsref}[1]{Sections~\ref{#1}}
\newcommand{\algthmref}[1]{Algorithm~\ref{#1}}
\newcommand{\algref}[1]{Algorithm~\ref{#1}}

\newcommand{\eref}[1]{Equation~\eqref{#1}}

\newcommand{\strassen}{\mbox{\sc Strassen}}

\newcommand{\gemm}{{\sc gemm}\xspace}

\newlength\savedwidth
\newcommand\whline{\noalign{\global\savedwidth\arrayrulewidth
                            \global\arrayrulewidth 1.5pt}%
           \hline
           \noalign{\global\arrayrulewidth\savedwidth}}

\newcommand*\mycircle[1]{\tikz[baseline=(char.base)]{
                    \node[shape=circle,draw,inner sep=1pt] (char) {#1};}}

\newcommand{\NoShow}[1]{}

\newcommand{\cutlass}{{\texttt{CUTLASS}}\xspace}

\newcommand{\cublas}{{\texttt{cuBLAS}}\xspace}

\newcommand*{\affaddr}[1]{#1} 
\newcommand*{\affmark}[1][*]{\textsuperscript{#1}}
\newcommand*{\email}[1]{\texttt{#1}}

\date{
August 23, 2018
}

\begin{document}

\title{
\LARGE Implementing Strassen's Algorithm with CUTLASS\\ on NVIDIA Volta GPUs
\\[0.2in]
\large FLAME Working Note \#88
}

\author{
Jianyu Huang\affmark[*]\affmark[\dag],
Chenhan D. Yu\affmark[*]\affmark[\dag],
Robert A. van de Geijn\affmark[*]\affmark[\dag]\\
\affaddr{\affmark[*]Department of Computer Science}\\
\affaddr{\affmark[\dag]Institute for Computational Engineering and Sciences}\\
\affaddr{The University of Texas at Austin, Austin, TX 78712}\\
\email{{\tt \{jianyu.huang@}, {\tt chenhan@}, {\tt rvdg@cs.\}utexas.edu}}
}



\maketitle

\begin{abstract}


Conventional GPU implementations of Strassen's algorithm (\strassen{}) typically rely on the existing high-performance matrix multiplication (\gemm{}),
trading space for time.
As a result, such approaches can only achieve practical
speedup for relatively large, ``squarish'' matrices due to the extra memory overhead, and their usages are limited due to the
considerable workspace.
We present novel \strassen{} primitives for GPUs that can be composed to generate a family of \strassen{} algorithms.
Our algorithms utilize both the memory and thread hierarchies on GPUs, reusing shared memory and register files inherited from \gemm{}, fusing additional operations, and avoiding extra workspace. We further exploit intra- and inter-kernel
parallelism by batching, streaming, and employing atomic operations.
We also develop a performance model for NVIDIA Volta GPUs to select the appropriate blocking parameters and predict the performance for \gemm{} and \strassen{}.
Overall, our 1-level \strassen{} can achieve up to $1.11\times$ speedup
with a crossover point as small as 1,536 compared to \texttt{cublasSgemm} on a NVIDIA Tesla V100 GPU. With additional
workspace, our 2-level \strassen{} can achieve $1.19\times$ speedup with a crossover point at 7,680.



\end{abstract}

\section{Introduction}
\label{s:gpu_intro}

Given matrices $A\in\mathcal{R}^{m\times k}$, $B\in\mathcal{R}^{k\times n}$,
and $C\in\mathcal{R}^{m\times n}$, Strassen's algorithm (\strassen{})~\cite{Strassen} 
computes matrix multiplication (\gemm{}
defined in \texttt{BLAS}~\cite{BLAS3} and \cublas{}~\cite{cublas})
\begin{equation} \label{eqn:prime}
	C = \alpha A \times B + \beta C
\end{equation}
with less than $\mathcal{O}(n^3)$ work.
The algorithm partitions the matrices into $2\times2$
submatrices such that
\begin{equation} \label{eqn:strassenpart}
\begin{bmatrix}
{C}_{0} & {C}_{1} \\
{C}_{2} & {C}_{3} \\
\end{bmatrix} = \alpha 
\begin{bmatrix}
{A}_{0} & {A}_{1} \\
{A}_{2} & {A}_{3} \\
\end{bmatrix} 
\begin{bmatrix}
{B}_{0} & {B}_{1} \\
{B}_{2} & {B}_{3} \\
\end{bmatrix} + \beta
\begin{bmatrix}
{C}_{0} & {C}_{1} \\
{C}_{2} & {C}_{3} \\
\end{bmatrix}, 
\end{equation}
and rearrange the arithmetic operations to reduce the number of
submatrix multiplications
from 8 to 7
(see~\secref{s:method} for details).
By recursively applying this scheme, it can be shown~\cite{Strassen} that
\eqref{eqn:prime} only requires $\mathcal{O}(n^{2.81})$ work.

\begin{figure}[tb!]
    \centering \includegraphics[scale=.55]{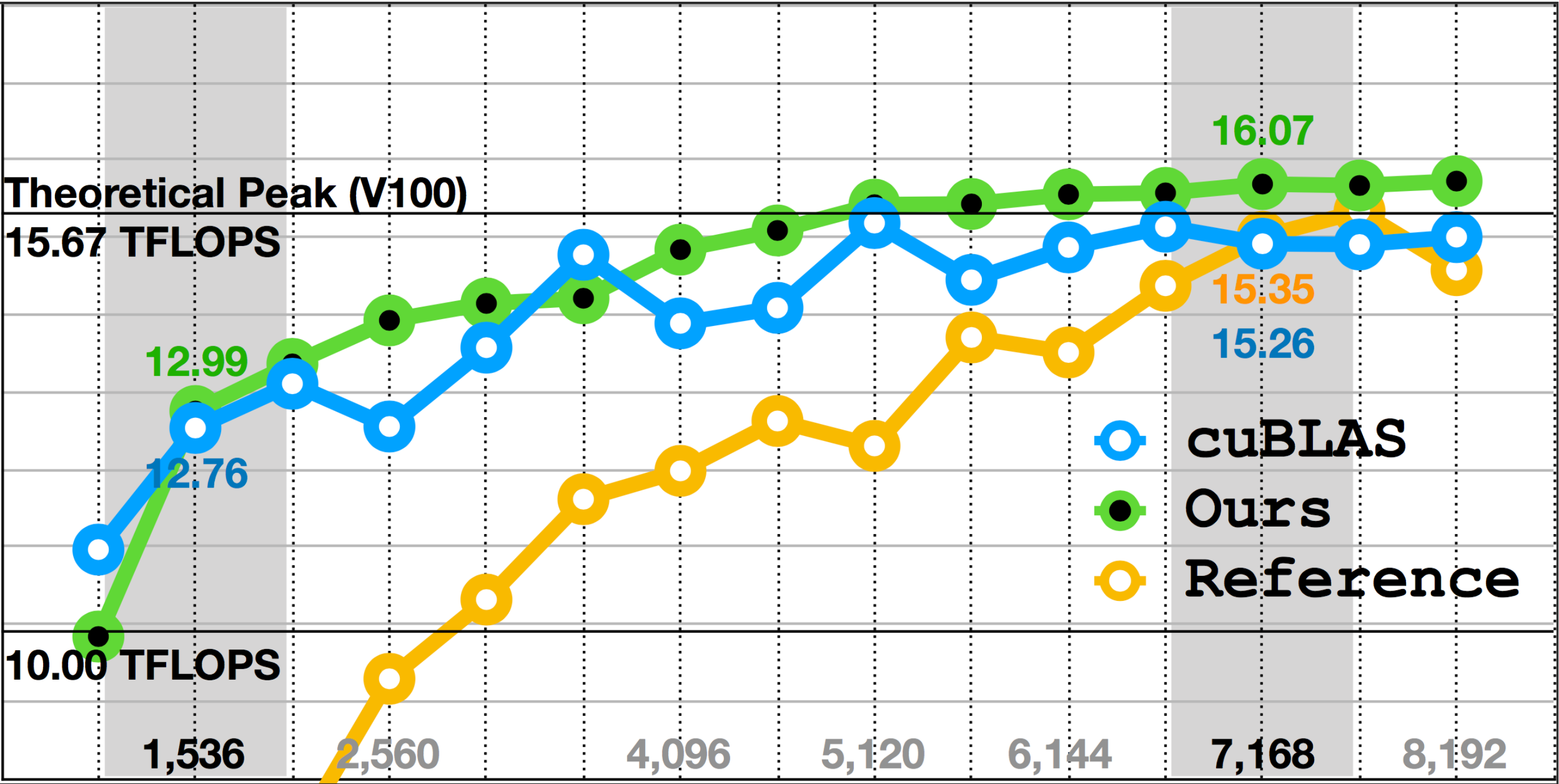} 
	\caption{\small Break-even point of our \strassen{} implementation
	and the state-of-the-art~\cite{StrassenGPU2}: the \textbf{x-axis} denotes the problem size
	($m=n=k$), and the \textbf{y-axis} denotes the floating point operation efficiency
	in \texttt{TFLOPS}.
	For a square matrix-multiplication,
	this work can achieve speedup over \texttt{cublasSgemm} for problem size
	as small as 1,536 while the state-of-the-art requires at least 7,168
	to break even (10k is required to obtain a stable speedup).
	}
  \label{fig:intro_comparison}
\end{figure}

Although it is easy to observe the saving from the complexity
analysis, the achievable practical speedup is typically disappointing
due to the extra memory overhead and space requirement~\cite{Benson15,D'alberto:2011:EPM:2049662.2049664,Strassen:SC16}
(see~\figref{fig:intro_comparison}).
A recent paper \cite{Strassen:SC16} addresses these issues and provides a good review on
the related work on modern CPU architectures. We  extend the idea in \cite{Strassen:SC16} and present a new \strassen{} algorithm on GPUs.

\noindent\textbf{Challenges:}
A practical \strassen{} implementation on GPUs must overcome several challenges.
First, the GPU architecture and programming model are different from 
their counterparts for a CPU. In order to achieve high performance, 
a practical implementation of \strassen{} needs to leverage the memory and thread hierarchies 
on GPUs.
Second, a GPU has a limited physical memory capacity. The conventional \strassen{} implementations require some extra temporary memory for storing intermediate submatrices, which limit the maximum problem size that can be computed compared to \gemm{} because of the GPU memory capacity. 
Third, a GPU is a highly parallel, multi-thread, many-core processor. 
\strassen{} needs to be parallelized at multiple granularities to fully utilize 
the computational horsepower of GPUs. 
There is thus a tension
between reducing the memory and exploiting more parallelism with the 
conventional implementation of \strassen{}.
Finally, the ratio between the theoretical peak performance and memory 
bandwidth of a GPU is even higher (less favorable) than that of a CPU. 
\strassen{} has a lower ratio of arithmetic operations to memory operations 
compared to \gemm{}, which means \strassen{} only becomes advantageous 
when the problem sizes are sufficiently large.
As a result, the practical implementation of \strassen{} needs to reduce the 
extra data movement to save the bandwidth and outperform \gemm{} for 
small or moderate problem sizes.

\noindent\textbf{Contributions:} 
Inspired by~\cite{Strassen:SC16} and the recent development of
\cutlass{}~\cite{cutlass}
(reviewed in~\secref{s:cutlass}), we introduce new algorithms for the practical implementation of
\strassen{} on GPUs. To be specific,  
\begin{itemize}[leftmargin=*]
	\item We develop new GPU \strassen{} kernels (\secref{s:gpu_stra}), which fuse additional memory and arithmetic operations with the \gemm{} pipeline.
	As a result, no additional workspace (GPU global memory and shared memory) is required.
	\item We present and discuss different optimization schemes and generate
	      different kernels that effectively reduce the number of required registers (\secref{s:register}).
	\item Our algorithms exploit both intra- and inter-kernel task-based parallelism.
				This allows us to maintain the parallelism without
				increasing the kernel launching and context switching overhead (\secref{s:task}).
	\item We derive an accurate performance model on NVIDIA Volta GPUs, which can help us to choose the right blocking parameters and predict the performance for \gemm{} and \strassen{} (\secref{s:analysis}).
	\item We conduct experiments on different matrix shapes
	(\secref{s:gpu_exp}).
	For square cases, our 1-level \strassen{} has a break-even point (faster than \texttt{cublasSgemm}) as small as 1,536, 
	while the state-of-the-art requires at
	least 7,168.
	Our hybrid 2-level \strassen{} has a break-even point as small as 7,680, while the state-of-the-art requires at least 13,312. Our implementations are also more efficient for non-square cases.
\end{itemize}
\noindent\textbf{Limitation:}
While the proposed approach does not require extra workspace (in the
global memory), it still trades memory operations 
(\texttt{mops}) for floating point operations (\texttt{flops}).
As a result, it may not be the optimal algorithm, and extra space
is preferred to offload the increasing register requirement and the global 
memory latency while applying \strassen{} algorithms to multiple levels.
This trade-off is discussed in \secref{s:analysis}.
Furthermore, \strassen{} is known to experience degradation in numerical stability, especially when more than two levels of recursions are incorporated \cite{HighamBook,DemmelStrassenStable2007,Ballard15}.
For this reason, only a few levels of recursions are leveraged.

\noindent\textbf{Related work:}
The literature on the theory and practice of \strassen{} is vast.
For a review, see~\cite{Strassen:SC16}. To our knowledge, there are no practical \strassen{} implementations on GPUs~\cite{StrassenGPU1,StrassenGPU2} that can be free from extra workspace and have a break-even point
as small as 1,536. The only algorithm and software that comes close 
is~\cite{StrassenGPU2}, which still requires additional $\mathcal{O}(mk/4+kn/4+mn/4)$ space.
In~\secref{s:gpu_exp}, we also provide empirical results with this algorithm as 
a reference. The idea of operation-fusing has also been generalized
to other domains to effectively reduce slow memory operations 
(improve temporal locality of the cache hierarchy) and extra
space requirement in tensor contraction and other $N$-body operations.
For a review, see~\cite{TC:Devin,Paul16,TensorStrassen,Chenhan:SC15}.
High-performance \gemm{} implementations on GPUs can be found in
\cite{GPUgemm1,GPUgemm2,GPUgemm3,NervanaGPU,GPUmodel1,GPUmodel2}.

\section{Background}
\label{s:gpu_background}
In this section, we first briefly review the 
GPU programming model (CUDA) and the GPU architecture (Volta)
we use in this paper. We then review the state-of-the-art algorithm
(the \cutlass{} framework) of high-performance \gemm{} on GPUs.


\subsection{GPU Programming Model}
\label{s:cuda_model}

\begin{figure}[tp!]
\center
\includegraphics[width=1.0\textwidth]{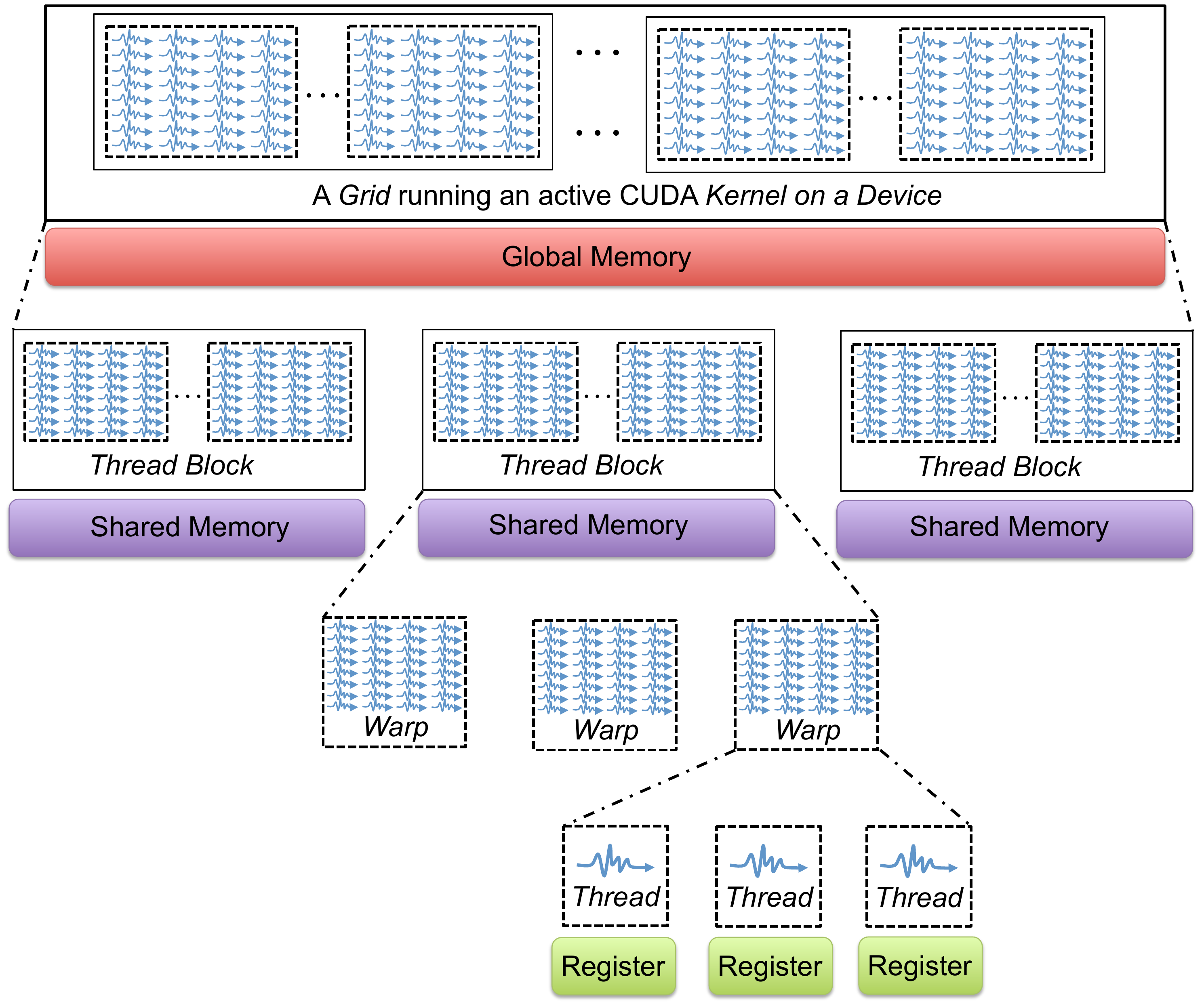}
\caption{The memory and thread hierarchies in the CUDA programming model.}
\label{fig:gpu_cuda_model}
\end{figure}

The CUDA programming model~\cite{cuda} assumes that the CUDA program (\emph{kernel}) is executed on physically independent \emph{devices} (GPUs) as coprocessors to the \emph{host} (CPU). \figref{fig:gpu_cuda_model} shows the memory and thread hierarchies on the GPU device.


\noindent\textbf{Memory hierarchy:}
The memory hierarchy on the GPU device includes three levels: global memory, shared memory (co-located with L1 and texture caches~\cite{volta3}), and register files.
The latency decreases while the bandwidth increases through the memory hierarchy from global memory to registers.


\noindent\textbf{Thread hierarchy:}
A \emph{thread} is the smallest execution unit in a 
CUDA program. 
A \emph{thread block} is a group of threads that run 
on the same core and 
shares a partition of resources such as shared memory.
Thread blocks communicate through barrier synchronization.
Multiple blocks are combined to form a \emph{grid}, which corresponds to 
an active CUDA kernel on the device.
At runtime, a thread block is divided into a number of \emph{warps} for 
execution on the cores. A warp is a set of 32 threads to execute the same 
instructions while operating on different data in lockstep.


\subsection{NVIDIA Volta GPUs} \label{s:volta}

We review the hardware specification of the NVIDIA Tesla V100~\cite{volta}, 
which features a GV100 (Volta) microarchitecture.
Tesla V100 is comprised of 80 streaming multiprocessors (SMs). Each SM is partitioned into 4 processing blocks. Each processing block consists of 2 Tensor Cores, 8 \texttt{FP64} (double precision) cores, 16 \texttt{FP32} (single precision) cores, and 16 \texttt{INT32} cores.
The tested Tesla V100 SXM2 GPU accelerator has the base clock frequency 1.3 GHz and boost clock frequency 1.53 GHz. As a result, the theoretical peak performance can reach 15.67 \texttt{TFLOPS}\footnote{1 \texttt{FMA}/cycle $\times$ 2 \texttt{flop/FMA} $\times$ 1.53G (boost clock frequency) $\times$ 16 (\# \texttt{FP32} core) $\times$ 4 (\# processing block/SM) $\times$ 80 (\# SM).} with single precision and 7.83 \texttt{TFLOPS}\footnote{1 \texttt{FMA}/cycle $\times$ 2 \texttt{flop/FMA} $\times$ 1.53G (boost clock frequency) $\times$ 8 (\# \texttt{FP64} core) $\times$ 4 (\# processing block/SM) $\times$ 80 (\# SM).} with double precision, while Tensor Cores can deliver 125 \texttt{TFLOPS}\footnote{64 \texttt{FMA}/cycle $\times$ 2 \texttt{flop/FMA} $\times$ 1.53G (boost clock frequency) $\times$ 2 (\# Tensor Core) $\times$ 4 (\# processing block/SM) $\times$ 80 (\# SM).} for \texttt{FP16/FP32} mixed precision.
The tested Tesla V100 GPU is built using 16 GB HBM2 memory with 900 GB/s of bandwidth.

\begin{figure*}
~
\begin{center}
\includegraphics[width=0.77\textwidth]{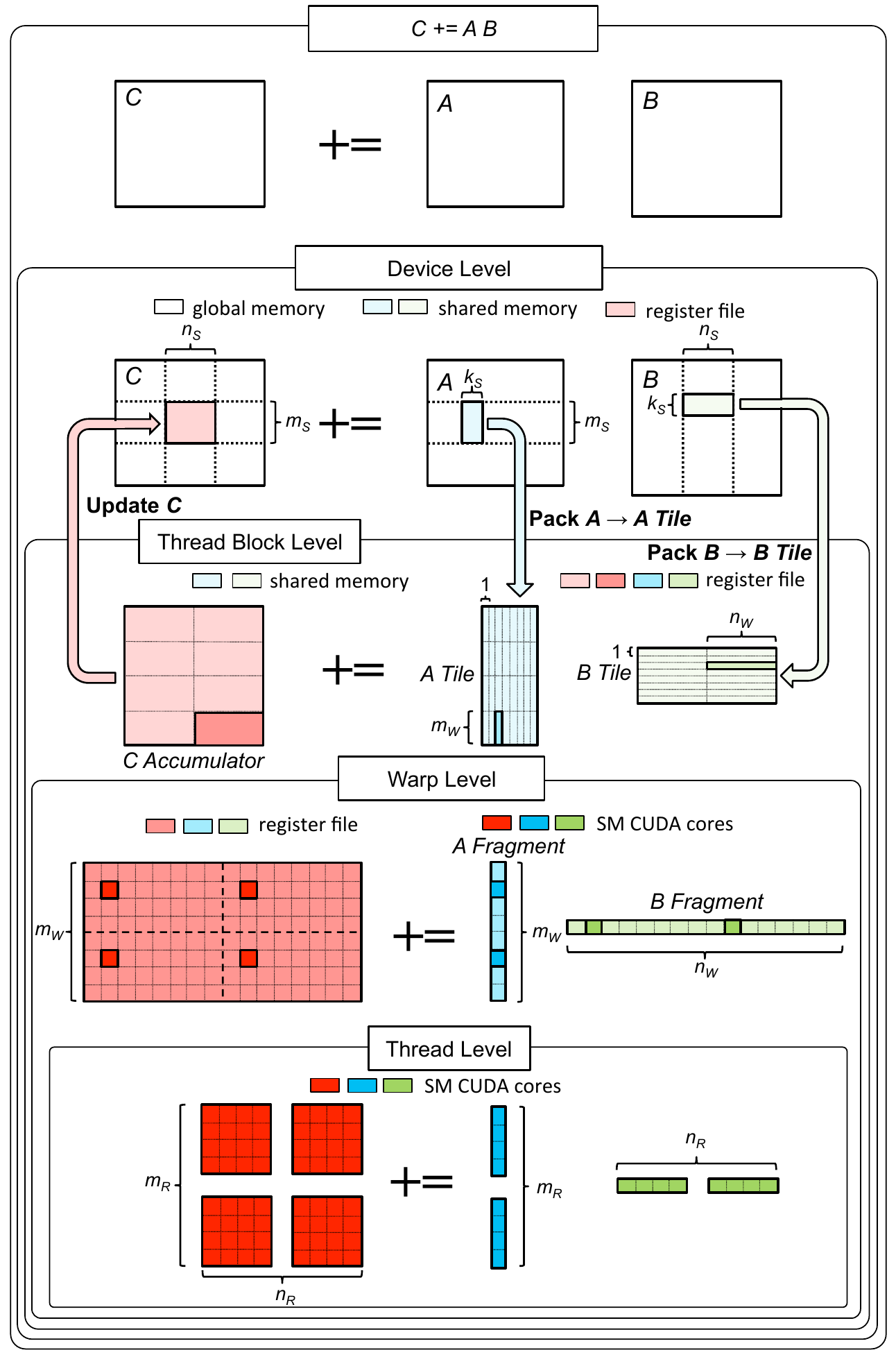}
\end{center}
\caption{
    Illustration of the \gemm{} implementation in CUTLASS~\cite{cutlass}.
    CUTLASS partitions the operand matrices into blocks in the different levels of the device, thread block, warp, and thread.
    Here we show block sizes typical for the large SGEMM:
$ m_S = 128 $, $n_S = 128$, $ k_S = 8 $;
$ m_W = 4 \times m_R = 32 $, $ n_W = 8 \times n_R = 64 $;
$ n_R = 8 $, $ n_R = 8 $.
}
\label{fig:gpu_gemm}
\end{figure*}

\subsection{Matrix Multiplication on GPUs} \label{s:cutlass}




We review the high-performance implementation of \gemm{} on NVIDIA GPUs, based on 
NVIDIA's CUDA Templates for Linear Algebra Subroutines (\cutlass{})~\cite{cutlass,cutlass2}, a collection of CUDA \texttt{C++} templates and abstractions to instantiate 
high-performance \gemm{} operations.
\cutlass{} incorporates strategies for hierarchical partition and 
data movement similar to \cublas{}~\cite{cublas}, the state-of-the-art 
implementation of the \texttt{BLAS} implementation on NVIDIA GPU, and can reach more than 90\% of \cublas{} performance on V100.
Without loss of generality, we will focus on single precision arithmetic and $ \alpha = \beta = 1 $ in \eqref{eqn:prime} henceforth.

\subsubsection{Blocking Strategies}
\label{s:gpu_blocking}

\figref{fig:gpu_gemm}
illustrates the \gemm{} implementation in \cutlass{}.
It organizes the computation by partitioning the operands into blocks in the different levels of the device, thread block, warp, and thread. 

\noindent\textbf{Device Level:} blocking for the thread blocks and shared memory.
The three operand matrices, $ A $, $ B $, and $ C $, are partitioned into $ m_S \times k_S $, $ k_S \times n_S $, and $ m_S \times n_S $ blocks.
Each thread block computes an $ m_S \times n_S $ block of $ C $ by accumulating the results of matrix products of an $ m_S \times k_S $ block of $ A $
and a $ k_S \times n_S $ block of $ B $.
Therefore, the $ m_S \times n_S $ block of $ C $ (the output of the thread block) is referred as the $ C $ \emph{Accumulator}. Since it is updated many times, it needs to be lifted into the fastest memory in the SM: the register files.
The global memory corresponding to the $ C $ \emph{Accumulator} only needs to be updated once after the $ C $ \emph{Accumulator} has accumulated the results of all matrix products along with the $ k $ dimension.
Furthermore, to improve data locality, blocks of $ A $ and $ B $ are ``\emph{packed}'' (copied) from global memory into shared memory (the $ A $ \emph{Tile} and $ B $ \emph{Tile}) for data reuse, accessible by all threads in the same thread block. 

\noindent\textbf{Thread Block Level:} blocking for the warps.
After the $ A $ \emph{Tile} and $ B $ \emph{Tile} are stored in shared memory, each individual warp computes a sequence of accumulated outer products by iteratively
loading an $ A $ \emph{Fragment}
(a subcolumn of the $ A $ \emph{Tile} with height $m_W$)
and a $ B $ \emph{Fragment}
(a subrow of the $ B $ \emph{Tile} with width $n_W$)
from the corresponding shared memory into register files along the $ k $ dimension and performing a rank-$1$ update.
The $ C $ \emph{Accumulator} is spatially partitioned across all the warps within the same thread block, with each warp storing a non-overlapping 2-D block in the register files.


\noindent\textbf{Warp Level:} blocking for the threads.
Each thread in a warp computes an $ m_R \times n_R $ outer product
with subvectors
of the $ A $ \emph{Fragment} and
subvectors
of the $ B $ \emph{Fragment} in a ``strip-mining'' (cyclic) pattern.
Each piece has a size of 4, because 
the largest granularity of vector load is 128 bits (4 single precision floating point numbers), and this helps to maximize the effective bandwidth. 
The total length of all pieces for an individual thread in $ m $ dimension is $ m_R $, while the total length in $ n $ dimension is $ n_R $.
Since each warp has 32 threads, 
\cutlass{} organizes the threads within the same warp in a $ 4 \times 8 $ or $ 8 \times 4 $ fashion such that
$ m_W / m_R = 4 $, $ n_W / n_R = 8 $, or $ m_W / m_R = 8 $, $ n_W / n_R = 4 $.

\noindent\textbf{Thread Level:} executing on the CUDA cores.
Each thread issues a sequence of independent FMA instructions to the CUDA cores and accumulates an $ m_R \times n_R $ outer product.

\subsubsection{Choices of Block Sizes}
\cutlass{} customizes six different strategies of block sizes
at each level $ \{ m_S$, $n_S$, $k_S$, $m_R$, $n_R$, $m_W$, $n_W \} $ in
\figref{fig:gpu_gemm}
for different matrix shapes and sizes, as shown in \tabref{tab:gpu_block_size}.
Details about how to choose these blocking parameters for large problem sizes are given in \secref{s:blocking_choose}.
Note that each thread block has
$ m_S / m_R \times n_S / n_R $ threads.



\begin{figure*}[!t]
\centering
{ 
\setlength{\tabcolsep}{15pt}
\begin{tabular}{ >{\columncolor[gray]{0.8}} c || r | r | r || r | r || c | c }
  \whline
  \rowcolor[gray]{0.8}
  Strategy & $ m_S $ & $ n_S $ & $ k_S $  &  $ m_R $ & $ n_R $ & $ m_W / m_R $ & $ n_W / n_R $ \\
\hline
  Small  &  16 &  16 & 16 & 2 & 2 & 4 & 8 \\
\hline                                    
  Medium &  32 &  32 &  8 & 4 & 4 & 4 & 8 \\
\hline                                    
  Large  &  64 &  64 &  8 & 8 & 8 & 4 & 8 \\
\hline                                    
  Tall   & 128 &  32 &  8 & 8 & 4 & 8 & 4 \\
\hline                                    
  Wide   &  32 & 128 &  8 & 4 & 8 & 4 & 8 \\
\hline                                    
  Huge   & 128 & 128 &  8 & 8 & 8 & 4 & 8 \\
  \whline
  \end{tabular}
}
\caption{
\texttt{CUTLASS} specifies six strategies of block sizes at each level in
\figref{fig:gpu_gemm}
for different matrix shapes and sizes.
}
\label{tab:gpu_block_size}
\end{figure*}


\subsubsection{Software Prefetching}

As shown in the left of \algthmref{a:gpu_stra}, to keep the SM busy, 
\cutlass{} uses global and local software prefetching to hide 
the data movement latency.
The computations on the CUDA cores are overlapped with the data 
preloading from the global memory (line~\ref{l:A_gr} and \ref{l:B_gr} in \algthmref{a:gpu_stra}) and from the shared memory (line~\ref{l:A_sr} and \ref{l:B_sr}).
A synchronization (line~\ref{l:sync}) is required to ensure that
all shared memory writes to
$tile_A$ and $tile_B$ between line~\ref{l:A_rs} and \ref{l:B_rs} have completed
before reading their values between line~\ref{l:A_gr} and \ref{l:B_gr} in the 
next iteration\footnote{\cutlass{} also provides the option of double buffering on the thread block level to enable concurrent reading for the current iteration and writing for the next iteration.
It eliminates the synchronization but also doubles the cost of the shared memory and the number of registers to hold the global memory fetches.
On the Tesla V100 GPUs, the option of double buffering on the thread block level is disabled.
}.

\afterpage{%
  \clearpage
\begin{landscape}
\begin{algorithm*}[t!]
\footnotesize
\begin{algorithmic}
  {
  \setlength\multicolsep{0pt}%
  \setlength{\columnseprule}{0.1pt}
	\begin{multicols}{2}
        \STATE\texttt{01:Register: $ frag_A[2][ m_R ], frag_B[2][ n_R ] $}
        \STATE\texttt{02:Register: $ next0_A[ m_R ], next0_B[ n_R ]$}
		\STATE\texttt{03:NOP}  \label{l:extra_reg1}
        \STATE\texttt{04:Register: $ accum_C[ m_R \times n_R ] $}
        \STATE\texttt{05:Shared memory: $ tile_A[ k_S \times m_S ], tile_B[ k_S \times n_S ] $}
		\STATE\texttt{06:load} one $ m_S \times k_S $ of $ A $  into $ tile_A[ k_S ][ m_S ] $ 
        \STATE\texttt{07:load} one $ k_S \times n_S $ of $ B $  into $ tile_B[ k_S ][ n_S ] $
		\STATE\texttt{08:\_\_syncthreads()}
		\STATE\texttt{09:load} subvectors of first column in $ tile_A $ into  $ frag_A[0][m_R] $
        \STATE\texttt{10:load} subvectors of first row in $ tile_B $ into  $ frag_B[0][n_R] $
        \STATE\texttt{11:}\texttt{\bf for} $ block\_k = 0 : k_S : k $ \texttt{\bf then}
        \STATE\texttt{12:}\gap\texttt{\textcolor{blue}{prefetch}} one subcolumn of next $m_S \times k_S$ block of $ A $ into $ next0_A[m_R] $ \label{l:A_gr}
		\STATE\texttt{13:}\gap\texttt{NOP}  \label{l:extra_mop1}
        \STATE\texttt{14:}\gap\texttt{\textcolor{blue}{prefetch}} one subrow of next $k_S \times n_S$ block of $ B $ into $ next0_B[n_R] $ \label{l:B_gr}
		\STATE\texttt{15:}\gap\texttt{NOP} \label{l:extra_mop2}
        \STATE\texttt{16:}\gap\texttt{\bf for} $ warp\_k = 0 : 1 : k_S $ \texttt{\bf then}
        \STATE\texttt{17:}\gap\gap\texttt{\textcolor{red}{prefetch}} next subcolumns in $ tile_A $ into  $ frag_A[(warp\_k+1)\%2][m_R] $ \label{l:A_sr}
        \STATE\texttt{18:}\gap\gap\texttt{\textcolor{red}{prefetch}} next subrows in $ tile_B $ into  $ frag_B[(warp\_k+1)\%2][n_R] $ \label{l:B_sr}
        \STATE\texttt{19:}\gap\gap$accum_C[m_R][n_R] \mathrel{+}=  frag_A[warp\_k\%2][m_R] frag_B[warp\_k\%2][n_R] $ \label{l:accum}
        
        \STATE\texttt{20:}\gap\texttt{store} $ next_A[ m_R ] $ into $ tile_A[ k_S ][ m_S ] $ \label{l:extra_flop1} \label{l:A_rs}
        \STATE\texttt{21:}\gap\texttt{store} $ next_B[ n_R ] $ into $ tile_B[ k_S ][ n_S ] $ \label{l:extra_flop2} \label{l:B_rs}
		\STATE\texttt{22:}\gap\texttt{\_\_syncthreads()} \label{l:sync}
        \STATE\texttt{23:}\texttt{write back} $ accum_C[m_R][n_R] $ to $ m_S \times n_S $ block of $ C $ \label{l:C_rg}
		\STATE\texttt{24:}\texttt{NOP} \label{l:extra_mop3}
        \STATE\texttt{01:Register: $ frag_A[2][ m_R ], frag_B[2][ n_R ] $}
        \STATE\texttt{02:Register: $ next0_A[ m_R ], next0_B[ n_R ]$}
		\STATE\texttt{03:Register: $ next1_A[ m_R ], next1_B[ n_R ] $}
        \STATE\texttt{04:Register: $ accum_C[ m_R \times n_R ] $}
        \STATE\texttt{05:Shared memory: $ tile_A[ k_S \times m_S ], tile_B[ k_S \times n_S ] $}
		\STATE\texttt{06:load} the sum of one $ m_S \times k_S $ of $ X $ and corresponding $ m_S \times k_S $ of $ Y $ into $ tile_A[ k_S ][ m_S ] $
        \STATE\texttt{07:load} the sum of one $ k_S \times n_S $ of $ V $ and corresponding $ k_S \times n_S $ of $ W $ into $ tile_B[ k_S ][ n_S ] $
		\STATE\texttt{08:\_\_syncthreads()}
		\STATE\texttt{09:load} subvectors of first column in $ tile_A $ into  $ frag_A[0][m_R] $
        \STATE\texttt{10:load} subvectors of first row in $ tile_B $ into  $ frag_B[0][n_R] $
        \STATE\texttt{11:}\texttt{\bf for} $ block\_k = 0 : k_S : k $ \texttt{\bf then}
		\STATE\texttt{12:}\gap\texttt{\textcolor{blue}{prefetch}} one subcolumn of next $m_S \times k_S$ block of $ X $ into $ next0_A[m_R] $
		\STATE\texttt{13:}\gap\texttt{\textcolor{blue}{($\delta$) prefetch}} one subcolumn of next $m_S \times k_S$ block of $ Y $ into $ next1_A[m_R] $
		\STATE\texttt{14:}\gap\texttt{\textcolor{blue}{prefetch}} one subrow of next $k_S \times n_S$ block of $ V $ into $ next0_B[n_R] $
		\STATE\texttt{15:}\gap\texttt{\textcolor{blue}{($\epsilon$) prefetch}} one subrow of next $k_S \times n_S$ block of $ W $ into $ next1_B[n_R] $
		\STATE\texttt{16:}\gap\texttt{\bf for} $ warp\_k = 0 : 1 : k_S $ \texttt{\bf then}
		\STATE\texttt{17:}\gap\gap\texttt{\textcolor{red}{prefetch}} next subcolumns in $ tile_A $ into  $ frag_A[(warp\_k+1)\%2][m_R] $
		\STATE\texttt{18:}\gap\gap\texttt{\textcolor{red}{prefetch}} next subrows in $ tile_B $ into  $ frag_B[(warp\_k+1)\%2][n_R] $
        \STATE\texttt{19:}\gap\gap$ accum_C[m_R][n_R] \mathrel{+}=  frag_A[warp\_k\%2][m_R] frag_B[warp\_k\%2][n_R] $
		\STATE\texttt{20:}\gap\texttt{($\delta$) store} $ next0_A[ m_R ] + next1_A[ m_R ] $ into $ tile_A[ k_S ][ m_S ] $
		\STATE\texttt{21:}\gap\texttt{($\epsilon$) store} $ next0_B[ n_R ] + next1_B[ n_R ] $ into $ tile_B[ k_S ][ n_S ] $ 
		\STATE\texttt{22:}\gap\texttt{\_\_syncthreads()}
		\STATE\texttt{23:}\texttt{write back} $ accum_C[m_R][n_R] $ to $ m_S \times n_S $ block of $ D $
		\STATE\texttt{24:}\texttt{($\gamma_1$) write back} $ accum_C[m_R][n_R] $ to $ m_S \times n_S $ block of $ E $
  \end{multicols}
  } 
\end{algorithmic}
\caption{ Comparisons between $ C +\!\!= A B $ and $ M = ( X + Y)( V+W); D +\!\!= M; E +\!\!= M $ on GPUs with software prefetching } 
\label{a:gpu_stra}
\end{algorithm*}
\end{landscape}
}

\section{Method} \label{s:method} Following~\cite{Strassen:SC16}, if the three operands $A$, $B$, and $C$ 
in~\eqref{eqn:prime} are partitioned into quadrants as in \eqref{eqn:strassenpart},
then 
\begin{equation}
{
\begin{array}{l @{\hspace{5pt}}  l @{\hspace{1pt}} c @{\hspace{1pt}} l l r}
    \mycircle{0} & {M}_0 &=&  ( {A}_{0} + {A}_{3} ) ( {B}_{0} + {B}_{3} );
&
{C}_{0} +\!\!= {M}_0;  {C}_{3} +\!\!= {M}_0;  \\
\mycircle{1} & {M}_1 &=&  ( {A}_{2} + {A}_{3} ) {B}_{0};
&
{C}_{2} +\!\!= {M}_1 ;  {C}_{3} -\!\!= {M}_1 ; \\
\mycircle{2} & {M}_2 &=&  {A}_{0} ( {B}_{1} - {B}_{3} );
&
{C}_{1} +\!\!= {M}_2 ;  {C}_{3} +\!\!= {M}_2 ;
\\
\mycircle{3} & {M}_3 &=&  {A}_{3}( {B}_{2} - {B}_{0} );
&
{C}_{0} +\!\!= {M}_3 ;  {C}_{2} +\!\!= {M}_3 ;
\\
\mycircle{4} & {M}_4 &=&  ( {A}_{0} + {A}_{1}) {B}_{3};
&
{C}_{1} +\!\!=  {M}_4 ;   {C}_{0} -\!\!= {M}_4;
\\
\mycircle{5} & {M}_5&=&  ({A}_{2} - {A}_{0} )( {B}_{0} + {B}_{1} );
&
{C}_{3} +\!\!= {M}_5;
\\
\mycircle{6} & {M}_6&=&  ({A}_{1} - {A}_{3} )( {B}_{2} + {B}_{3} );
&
{C}_{0} +\!\!= {M}_6 ;
\end{array}
}
\label{eqn:gpu_allops}
\end{equation} %
compute $ {C} \,+\!\!= {A B} $, with seven instead of eight (sub)matrix multiplications,
decreasing the total number of arithmetic operations
by a factor of $ 7 / 8 $ (ignoring total number of extra additions, a lower order term).
If all matrices are square and of size $ N \times N $,
theoretically this single step of \strassen{}~\cite{Strassen} can be applied 
recursively, resulting in the classical \strassen\ with a cost of $ \mathcal{O} ( N^{2.81} ) $.

The operations above in~(\ref{eqn:gpu_allops}) are all 
instances of the following general \strassen{} primitive
\begin{equation}
{M} = ( {X} + \delta {Y})( {V} + \epsilon {W}); ~~~{D} +\!\!= \gamma_0 {M}; ~~~{E}+\!\!= \gamma_1 {M};
\label{e:straprim1}
\end{equation}
with $ \gamma_0, \gamma_1, \delta, \epsilon \in \{ -1, 0, 1 \} $.
Here, ${X}$ and ${Y}$ are submatrices of ${A}$, ${V}$ and ${W}$ are 
submatrices of ${B}$, and ${D}$ and ${E}$ are submatrices of ${C}$.
This scheme can be extended to
multiple levels of \strassen{}~\cite{Strassen:SC16}.

We present a new GPU kernel that computes~(\ref{e:straprim1}) in~\secref{s:gpu_stra}. 
We discuss how to effectively reduce the register requirement
and generate different kernel variants in~\secref{s:register}.
Task parallelism is discussed in~\secref{s:task}. 
Two-level \strassen{} algorithms and fringe case handling
are discussed in~\secsref{s:2level} and~\ref{s:fringe}.

\subsection{Strassen's Algorithm on NVIDIA GPUs} \label{s:gpu_stra} 
\afterpage{%
\clearpage
\begin{figure*}[tp!]
~
\begin{center}
\includegraphics[width=0.77\textwidth]{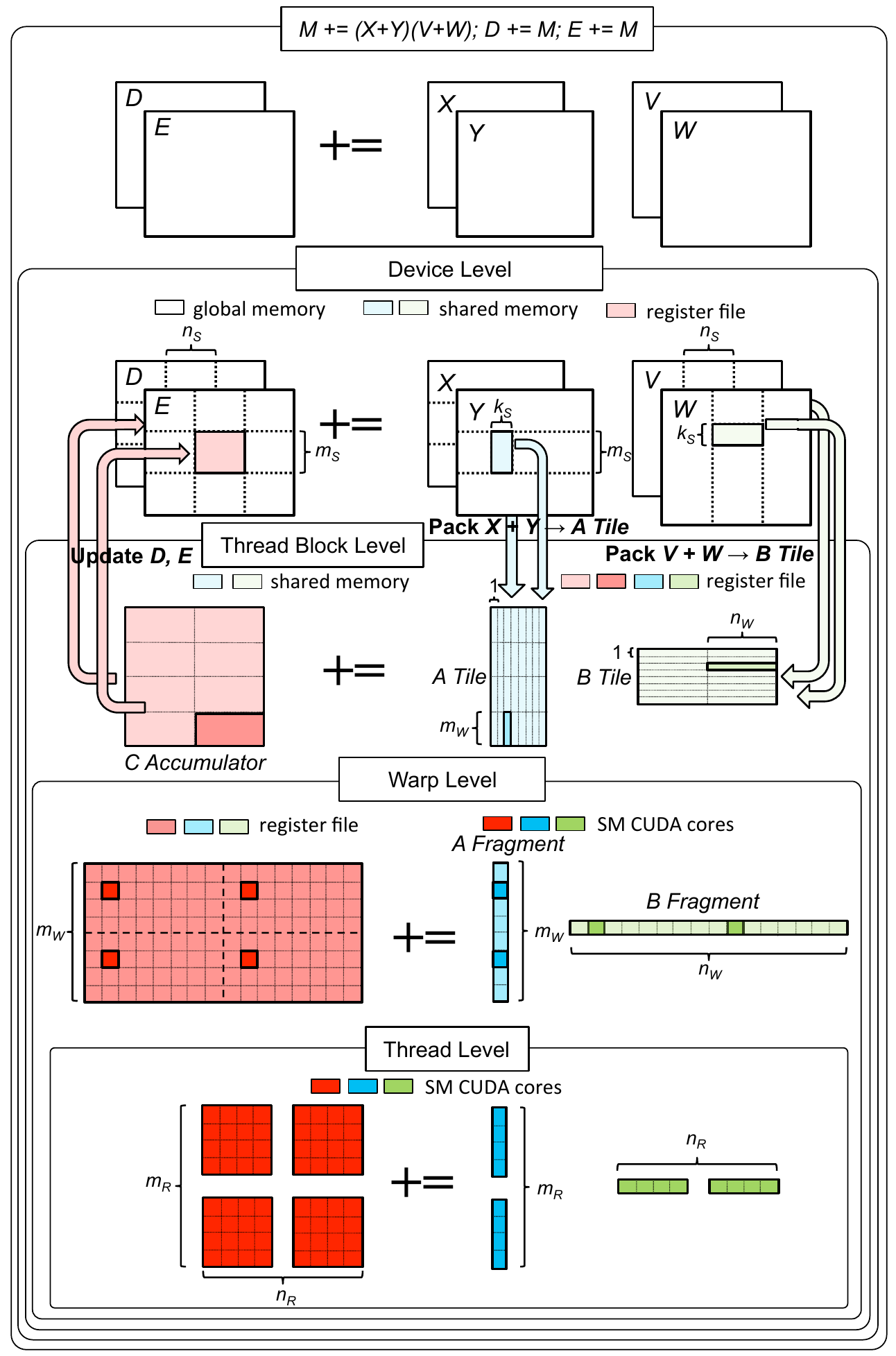}
\end{center}
\caption{
    Specialized kernel that implements the representative computation $ M = ( X + Y)( V+W); D +\!\!= M; E +\!\!= M $ of each row of computations
    in~\eqref{eqn:gpu_allops}
    based on \figref{fig:gpu_gemm}.
$ X $, $ Y $ are submatrices of $ A $; $ V $, $ W $ are submatrices of $ B $; $ D $, $ E $ are submatrices of $ C $; $ M $ is the intermediate matrix product.
}
\label{fig:gpu_stra}
\end{figure*}
}

\afterpage{%
  \clearpage
\begin{landscape}
\begin{figure*}
~
\begin{center}
\includegraphics[width=1.20\textwidth]{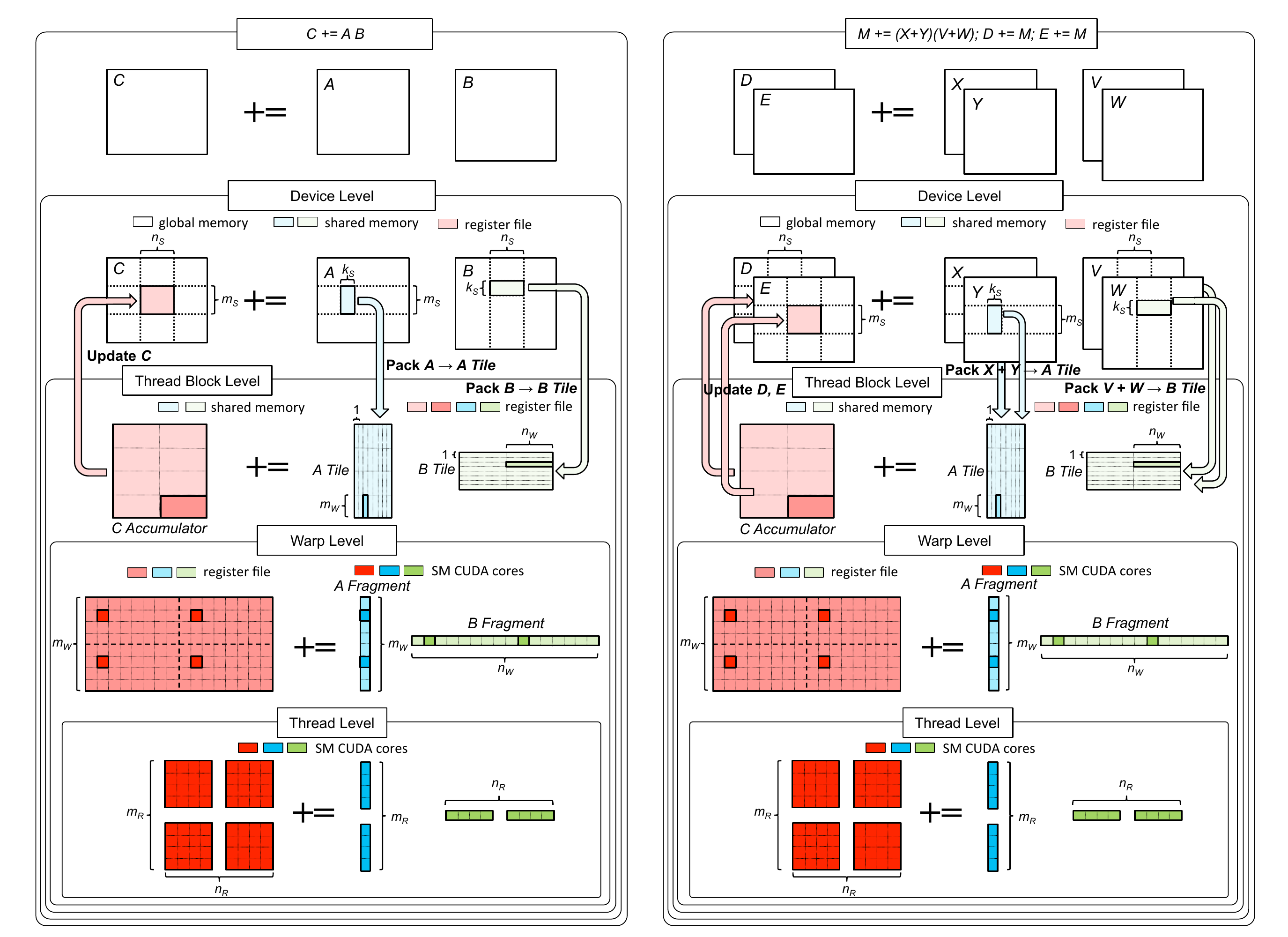}
\end{center}
\caption{
    A side-by-side comparison of the \gemm{} implementation in \cutlass{} and
    our modifications for implementing the representive computation
    $ M = ( X + Y)( V+W); D +\!\!= M; E +\!\!= M $. Left: \figref{fig:gpu_gemm}; Right: \figref{fig:gpu_stra}.
}
\label{fig:gpu_side_by_side}
\end{figure*}
\end{landscape}
}





We extend the \gemm{} implementation for GPUs illustrated in
\figref{fig:gpu_gemm}
to accommodate the \strassen{} primitive
\begin{equation}
 M = P Q = ( X + Y )( V + W ); D +\!\!= M; E +\!\!= M. 
\label{e:gpu_straprim2}
\end{equation}
The conventional approach
performs pre-processing 
on the inputs $P = (X+Y)$, $Q=(V+W)$, 
and post-processing on its outputs $D+\!\!=M$ and $E+\!\!=M$.
In other words, \emph{the conventional approach must introduce extra workspace 
(in the global memory) and memory operations for intermediate 
matrices $P$, $Q$, and $M$ to cast
(\ref{e:gpu_straprim2}) in terms of calls to 
\gemm{}.}

Instead of casting the primitive in terms of \gemm{}, we develop a 
specialized kernel utilizing the memory and thread hierarchies on GPUs and show how these pre-processing
and post-processing phases can be efficiently incorporated without
introducing extra workspace.
We illustrate how these extra memory operations (and a few floating
point operations) are fused in
\figref{fig:gpu_stra}
and \figref{fig:gpu_side_by_side} (right)
without affecting
the implementations in the warp and thread level:

\noindent\textbf{Packing the $A$ and $B$ \emph{Tiles}:}
The summation of matrices $ X + Y $ can be incorporated 
into the packed $ A $ \emph{Tile} during the packing process 
(from the \textbf{Device Level} to the \textbf{Thread Block Level} 
in \figref{fig:gpu_side_by_side}), avoiding the extra workspace requirement 
and reducing the additional memory movement since the $ A $ \emph{Tile} 
is reused for the temporary matrix sum, which is held in the shared memory.
Similarly, the summation of matrices $ V + W $ can be also 
incorporated into the packed $ B $ \emph{Tile} during the packing process.

\noindent\textbf{Writing back the $C$ $\emph{Accumulator}$:}
After the $ C $ \emph{Accumulator} has accumulated its result of 
$ ( X + Y ) ( V + W ) $ along the $ k $ dimension, it can update the 
appropriate parts of $ D $ and $ E $ in the global memory once 
(from the \textbf{Thread Block Level} to the \textbf{Device Level}). 
This optimization avoids the required workspace for intermediate 
matrices $ M_i $ and reduces the additional memory movement since the 
$ C $ \emph{Accumulator} is kept in the register files:
it is fetched from the global memory into the register once in the beginning,
and it is written to $ D $ and $ E $ only after its computation completes.

\subsection{Register Optimization} \label{s:register} \begin{figure}[!t]
    \centering
    {
        \setlength{\tabcolsep}{11pt}
    \begin{tabular}{ >{\columncolor[gray]{0.8}}c || c || p{0.1cm} | p{0.1cm} | p{0.1cm} | p{0.1cm} || p{0.1cm} | p{0.1cm} | p{0.1cm} | p{0.1cm} | p{0.1cm} | p{0.1cm} | p{0.1cm} | p{0.1cm} | p{0.1cm} | p{0.1cm} }
            \whline
            \rowcolor[gray]{0.8}
            &
            & \multicolumn{4}{c ||}{ 1-level } & \multicolumn{10}{c}{ 2-level } \\ 
            \hhline{~||~||----|----------}
             \rowcolor[gray]{0.8}
             Var\# & * & 0 & 1 & 2 & 3 & 0 & 1 & 2 & 3 & 4 & 5 & 6 & 7 & 8 & 9 \\
            \whline
            $ W_A $ & 1 & 2 & 2 & 1 & 2 & 4 & 1 & 2 & 4 & 4 & 4 & 4 & 2 & 2 & 4 \\
            $ W_B $ & 1 & 2 & 2 & 2 & 1 & 4 & 4 & 4 & 1 & 2 & 4 & 4 & 2 & 4 & 2 \\
            $ W_C $ & 1 & 2 & 1 & 2 & 2 & 4 & 4 & 4 & 4 & 4 & 1 & 2 & 4 & 2 & 2 \\
            \hline
            Cnt\#   & 1 & 1 & 2 & 2 & 2 & 1 & 4 & 4 & 4 & 4 & 4 & 4 & 8 & 8 & 8 \\
            \whline
        \end{tabular}
    }
    \caption{
    Operand and instance counts of \gemm{}, 1-level, and 2-level 
    \strassen{} primitives. The stared (*) column denotes the
    base case \gemm{}, which only has one operand per matrix.
    1-level \strassen{} primitives have at most two operands
    per matrix, and overall there are 7 instances with 4 different variants.
    2-level  \strassen{} primitives have at most four operands
    per matrix, and overall there are 49 instances with 10 different variants.
    }
    \label{tab:perfeq_level2}
\end{figure}

We give the implementation of the \strassen{} 
primitive (on the right) side-by-side with \texttt{CUTLASS}'s \gemm{} algorithm 
(on the left) in~\algref{a:gpu_stra}.
Recall that the primitive incorporates pre-processing and
post-processing steps to create a new kernel that avoids 
additional workspace.
As a result, we must (for the current NVIDIA GPU architecture) 
introduce extra registers at line~\ref{l:extra_reg1}, extra \texttt{mops}
at line~\ref{l:extra_mop1}, line~\ref{l:extra_mop2}, line~\ref{l:extra_mop3},
and extra \texttt{flops} at line~\ref{l:extra_flop1} and
line~\ref{l:extra_flop2}.

Notice that the algorithm presented in~\algref{a:gpu_stra} is the
 general form of the seven instances in~(\ref{eqn:gpu_allops}).
Depending on the value of scalars $\delta$, $\epsilon$ and $\gamma_1$ 
(represented as statement predicates in~\algref{a:gpu_stra}), 
we can generate specialized kernels at compile time (using \texttt{C++} non-type template parameters) that 
may optimize out these extra registers, \texttt{mops} and \texttt{flops}.
Overall, there are four different variants (\textbf{Var\#0}--\textbf{Var\#3}) for
the one-level \strassen{} and ten different variants
for the two-level \strassen{}.
These variants have different operand counts $W_{\{A,B,C\}}$, as shown in \tabref{tab:perfeq_level2}.

\noindent\textbf{Var\#0 and Var\#1:}
Instance \mycircle{0} in~\eqref{eqn:gpu_allops}, whose predicates are all
\texttt{true} (non-zero), forms Var\#0.
That is,  the instance in Var\#0, with the operand counts $ W_A = W_B = W_C = 2 $, contains additional register allocation at line~\ref{l:extra_reg1}, 
additional \texttt{mops} at line~\ref{l:extra_mop1},~~\ref{l:extra_mop2}, and \ref{l:extra_mop3}, as well as additional \texttt{flops} at line~\ref{l:extra_flop1} and \ref{l:extra_flop2}.
Var\#1 (with scalar $\gamma_1=0$) contains Instances \mycircle{5} and \mycircle{6}.
As a result, instances in Var\#1,  with the operand counts $ W_A = W_B = 2 $ and $ W_C = 1 $, do not perform
extra post-processing on the output, hence with fewer \texttt{mops}.
Both variants allocate registers $next1_{A}[m_R]$ and $next1_{B}[n_R]$, 
consuming the most registers out of the four variants.

\noindent\textbf{Var\#2 and Var\#3:}
Instances \mycircle{2} and \mycircle{3}, whose predicate $\delta$ is \texttt{false}, 
form Var\#2, with the operand counts $ W_A = 1 $ and $ W_B = W_C = 2 $. Because scalar $\delta = 0$, only registers $next1_A[m_R]$ will be 
allocated. Registers $next1_A[m_R]$ will be optimized out (through dead code elimination), since they will never be used.
Similarly, Instances \mycircle{1} and \mycircle{4} form Var\#3, 
	which has the operand counts $ W_B = 1 $ and $ W_A = W_C = 2 $ and only allocates registers $next1_B[n_R]$.
These two variants have smaller register pressure, typically 
performing slightly better (with higher \texttt{FLOPS}) than Var\#0 and
Var\#1 when the problem sizes are large.
See~\secref{s:analysis} for a quantitative analysis on how these variants
affect the performance.





\subsection{Task Parallelism} \label{s:task} A straightforward implementation of \Strassen{} based on 
our specialized kernel (\secref{s:gpu_stra}) invokes a sequence of 
GPU kernels sequentially (7 kernels for 1-level, 49 kernels for 2-level).
This approach achieves intra-kernel parallelism across 
the thread blocks, warps, and threads,
which is utilized in the \gemm{} implementation on a GPU.
However, it is further possible to improve concurrency by
exploiting more inter-kernel parallelism.

A careful look at (\ref{eqn:gpu_allops})
reveals that (i) the ordering of these operations can be arbitrary;
(ii) the dependencies between the kernels for these operations only occur 
for the concurrent writes to different submatrices of $ C $.
That is, as long as race conditions are resolved, we can compute
several instances in (\ref{eqn:gpu_allops}) simultaneously.
Inter-kernel parallelism is especially important for
small problem sizes when there is limited intra-kernel parallelism such that each kernel cannot saturate the workload on the GPU device
and
for multi-level \strassen{} when the partitioned block sizes are small.
We next present three schemes to achieve this goal.

\noindent\textbf{Streaming with dependencies:}
By invoking multiple independent kernels without write dependencies 
to different parts of $ C $, we can achieve inter-kernel parallelism.
To be specific, the seven instances in~(\ref{eqn:gpu_allops})
can be rearranged into three synchronous stages (Stage 0--2) according to the 
dependency analysis, where kernels in the
same stage can be executed asynchronously with two CUDA
\texttt{streams}\footnote{
CUDA programs can manage the concurrency across kernels through 
\emph{streams}~\cite{cuda}, each of which is a sequence of commands 
that execute in order.
While the kernels launched within the same stream must be scheduled 
in sequential order, the commands from different streams may run 
concurrently out of order. To ensure every command in a particular 
stream has finished execution, {\tt cudaDeviceSynchronize} can be used 
to enforce synchronization points.} (\texttt{stream[0]} and \texttt{stream[1]}).

\begin{figure*}
\center
{
\setlength{\tabcolsep}{18pt}
\begin{tabular}{ >{\columncolor[gray]{0.8}}c  |  l @{\hspace{5pt}}  l @{\hspace{1pt}} c @{\hspace{1pt}} l l @{\hspace{2pt}} | c }
    \whline
    \rowcolor[gray]{0.8}
    Stage & \multicolumn{5}{ c| }{Operation} & \texttt{stream} \\
    \hline
    $ 0 $ & \mycircle{1} & $ {M}_1 $ & $ = $ &  $ ( {A}_{2} + {A}_{3} ) {B}_{0};              $  & $ {C}_{2} +\!\!= {M}_1;  {C}_{3} -\!\!= {M}_1; $ & $ [0] $ \\
          & \mycircle{4} & $ {M}_4 $ & $ = $ &  $ ( {A}_{0} + {A}_{1}) {B}_{3};               $  & $ {C}_{1} +\!\!= {M}_4;  {C}_{0} -\!\!= {M}_4; $ & $ [1] $ \\
          & \mycircle{5} & $ {M}_5 $ & $ = $ &  $ ({A}_{2} - {A}_{0} )( {B}_{0} + {B}_{1} );  $  & $ {C}_{3} +\!\!= {M}_5;                        $ & $ [0] $ \\
          & \mycircle{6} & $ {M}_6 $ & $ = $ &  $ ({A}_{1} - {A}_{3} )( {B}_{2} + {B}_{3} );  $  & $ {C}_{0} +\!\!= {M}_6;                        $ & $ [1] $ \\
    \hline
    $ 1 $ & \mycircle{2} & $ {M}_2 $ & $ = $ &  $ {A}_{0} ( {B}_{1} - {B}_{3} );              $  & $ {C}_{1} +\!\!= {M}_2;  {C}_{3} +\!\!= {M}_2; $ & $ [0] $ \\
          & \mycircle{3} & $ {M}_3 $ & $ = $ &  $ {A}_{3}( {B}_{2} - {B}_{0} );               $  & $ {C}_{0} +\!\!= {M}_3;  {C}_{2} +\!\!= {M}_3; $ & $ [1] $ \\
    \hline
    $ 2 $ & \mycircle{0} & $ {M}_0 $ & $ = $ &  $ ( {A}_{0} + {A}_{3} ) ( {B}_{0} + {B}_{3} );$  & $ {C}_{0} +\!\!= {M}_0;  {C}_{3} +\!\!= {M}_0; $ & $ [0] $ \\
    \whline
\end{tabular}
}
\caption{
    Reordered operations based on~\eqref{eqn:gpu_allops} with multi-kernel streaming.
}
\label{tab:gpu_streaming}
\end{figure*} %



In~\tabref{tab:gpu_streaming}, Stage 0 contains four instances. 
Instances \mycircle{1} and \mycircle{4} can be executed concurrently with 
\texttt{stream[0]} and \texttt{stream[1]}. 
Instance \mycircle{5} can be executed right after \mycircle{1} using \texttt{stream[0]}
to avoid the possible race condition, and \mycircle{6} can be executed 
using \texttt{stream[1]} in the same way.
Instances \mycircle{2} and \mycircle{3} are executed concurrently in Stage 2,
and Stage 3 only contains Instance \mycircle{0}.
Both streams must be synchronized at the end of each stage
to enforce the order.

\noindent\textbf{Element-wise atomic update:}
Although the first scheme works reasonably well for large problem sizes
(where inter-kernel parallelism is less crucial), two streams do not expose enough parallelism for small and medium problem 
sizes (say $m = n = k \leq 6000$).
Instead of resolving the race condition in the granularity of
kernels, we exploit \emph{out-of-order} parallelism 
at a finer granularity using atomic operations to resolve
the possible concurrent write conflicts on matrix $C$.
This is done by replacing the normal \texttt{Add} in the 
\emph{Accumulator} with a global \texttt{atomicAdd} instruction.
As a result, the 7 instances can all be executed concurrently
with up to 7 CUDA \texttt{streams}.

\noindent\textbf{Batching:}
With \texttt{atomicAdd}, 
1-level \strassen{} launches 7 kernels concurrently, and 2-level
\strassen{} may launch up to 49 kernels simultaneously. 
Although multiple streams can introduce more parallelism, the performance can easily be compromised by the kernel launching and context switch overhead, which is proportional to the number of streams and kernels.
The overhead can even slow down the overall runtime when the problem size is small.
As a result, we seek to launch the minimum number
of kernels and streams by batching instances according to their variants.
Instances in the same variant can be realized as a sequence (batch) of
independent \strassen{} primitives (given the race condition 
on $C$ is resolved by \texttt{adtomicAdd}).

To be specific, we use four streams to launch four GPU kernels
concurrently.
For example, the two instances in Var\#1 are grouped as a batch of two, and the kernel is launched with 3D-grid, where the
\emph{z-dimension} equals the batch size.
\texttt{blockIdx.x} and \texttt{blockIdx.y} are used to create
the 2D-thread-block as usual to exploit parallelism
within each \strassen{} instance. The additional \texttt{blockIdx.z} 
is used as an offset to exploit task-based parallelism
between \strassen{} instances and access the proper pointers
and scalars toward $X$, $Y$, $V$, $W$, $D$, $E$, $\delta$, $\epsilon$,
$\gamma_0$, and $\gamma_1$.

\subsection{Two-Level Strassen's Algorithm} \label{s:2level} \noindent\textbf{Direct 2-level \strassen{}:}
Following~\cite{Strassen:SC16}, we can derive 49 instances (10 variants)
from the general 2-level \strassen{} primitive that resembles~\eqref{e:straprim1}
but with up to four submatrix operations in each operand. 
In the hierarchical view of
\figref{fig:gpu_stra},
we need to load four submatrices while packing the $A$ and $B$ \emph{Tiles}
from the \textbf{Device Level}.
We also need to write the output back to four submatrices
from the \textbf{Thread Block Level}. 
In~\algref{a:gpu_stra}, we need to allocate extra register blocks 
$next2_{A}[m_R]$, $next2_{B}[n_R]$, $next3_{A}[m_R]$, and $next3_{B}[n_R]$
at line~\ref{l:extra_reg1}.
Additional \texttt{mops} are introduced at line~\ref{l:extra_mop1}, \ref{l:extra_mop2}, and \ref{l:extra_mop3}. 
There are also additional \texttt{flops} introduced at 
line~\ref{l:extra_flop1} and \ref{l:extra_flop2}.
As we can observe, although implementing a 2-level \strassen{} primitive
can get rid of extra space requirement, the trade-off (regarding the
current NVIDIA GPU architecture) is to increase
the register pressure and the required memory bandwidth.
As a result, the occupancy and floating point operation efficiency
may be compromised. 
For a discussion on how this can be resolved in the future,
see~\secref{s:perf_discuss}.

\noindent\textbf{Hybrid 2-level \strassen{}:}
Alternatively, we combine the reference approach~\cite{StrassenGPU2} with 
our specialized kernel to relieve the register pressure and the
required memory bandwidth. 
The idea is to first apply the reference approach in~\cite{StrassenGPU2},
which requires $\mathcal{O}(mk/4+kn/4+mn/4)$ workspace.
Then we apply our 1-level \strassen{} primitive to each of the
seven submatrix multiplications.
Together, we have a hybrid 2-level \strassen{} algorithm that
consumes the same amount of workspace as~\cite{StrassenGPU2} but ramps up
much faster with smaller problem sizes.
We empirically compare our hybrid approach with~\cite{StrassenGPU2} in~\secref{s:gpu_exp}.

\subsection{Handling the Fringes} \label{s:fringe} Traditionally, for matrices with odd dimensions, we need to handle the remaining fringes before applying \strassen{}.
There are some well-known approaches such as padding (i.e., adding rows or columns with zeros to get matrices of even dimensions) and
peeling (i.e., deleting rows or columns to obtain even dimensioned matrices)~\cite{Huss-Lederman:1996:ISA:369028.369096, StrassenDynamicPeeling}
followed by post-processing.
In our approach, fringes can be internally handled by padding the $ A $ \emph{Tile} and $ B $ \emph{Tile} with zeros,
and aligning the $ m_C \times n_C $ $ C $ \emph{Accumulator} along the fringes.
This trick avoids the handling of the fringes with extra memory or computations because the packing and accumulation processes always occur for
the high-performance implementation of \gemm{} on GPUs, and we reuse the same buffers.

\section{Experiment}
\label{s:gpu_exp}
We conduct three sets of experiments in~\figref{fig:stra_square}, providing
an overview of our 1-level and 2-level \strassen{}.
We discuss and analyze the performance of our algorithms through modeling
in~\secref{s:analysis}.

\noindent\textbf{Setup:} We perform our experiments on a Tesla V100 SXM2 accelerator which is connected to an Intel Xeon Gold 6132 Skylake server.
The Operating System is CentOS Linux version 7.4.1708.
The GNU compiler version for compiling the host code is 6.4.0.
We use CUDA Toolkit 9.1 and compile the code with flags  {\tt -O3 -Xptxas -v -std=c++11 -gencode arch=compute\_70,code=sm\_70}.
As presented in \secref{s:volta}, the tested Tesla V100 SXM2 accelerator has a theoretical peak performance of 15.67 \texttt{TFLOPS} in single precision.

\noindent\textbf{Measurement:}
We report the single precision floating point efficiency with three 
different configurations in~\figref{fig:stra_square}.
We fix the ratio of $m$, $n$, and $k$ dimension in the first configuration
such that all matrices are square.
In the second configuration, we fix $k=4,096$ and vary $m$, $n$, resulting
in tall-and-skinny matrix-multiplication (rank-$k$ update).
In the last configuration, we fix $m=n=8,192$ and vary $k$, resulting
in short-and-fat matrix-multiplication (panel dot-product).

To measure the execution time of GPU kernels running, we use CUDA events that 
have a resolution of approximately half a microsecond.
We take \emph{Effective} \texttt{TFLOPS} as the main metric to compare the performance of various implementations. To be specific, 
\begin{equation}
  \text{\emph{Effective} \texttt{TFLOPS}} = \frac{2\cdot m\cdot n\cdot k}{\text{time (in seconds)}}\cdot 10 ^{-12}.
  \label{e:gpu_etflops}
\end{equation}
\cutlass{} and our methods are tested with different strategies 
and block sizes to select
the highest performing setup.


\begin{figure}[htp!]
\center
\begin{tikzpicture}[scale=1.05]
\begin{axis}[
    xlabel={ $m = k = n$ },
    x label style={at={(axis description cs:0.5,0.02)}},
    ylabel={\emph{Effective} TFLOPS},
    y label style={at={(axis description cs:0.06,0.5)}},
    xmin=0,
    xmax=20000,
    ymin=0,
    ymax=18.5,
    xtick={1000,2000,3000,4000,5000,6000,7000,8000,9000,10000,12000,14000,16000,18000,20000},
    ytick={2,4,6,8,10,12,14,15.667,18.5},
    scaled x ticks=false,
    scaled x ticks=base 10:-3,
    grid=major,
    axis background/.style={fill=lightgray!20},
    mark size=1pt,
    cycle list name=jianyugpucolor,
    restrict y to domain=1:inf,
    legend style={
        at={(0.99,0.01)},
        anchor=south east,
        legend columns=2,
        font=\tiny,
        rounded corners=1pt,
        nodes={scale=1.1, transform shape},
        cells={anchor=west},
    },
    legend entries = {
        \cutlass{} \\
        \cublas{} \\
        $ 1~level, Ours $\\
        $ 1~level, Reference $\\
        $ 2~level, Hybrid $\\
        $ 2~level, Reference $\\
    },
    ]

\addplot table[x=dim,y=cutlass,col sep=comma]          {plotdata/stra_square_V100_SXM2_final.csv};
\addplot table[x=dim,y=cublas,col sep=comma]           {plotdata/stra_square_V100_SXM2_final.csv};
\addplot table[x=dim,y=one_abc_opt,col sep=comma]      {plotdata/stra_square_V100_SXM2_final.csv};
\addplot table[x=dim,y=one_ref,col sep=comma]          {plotdata/stra_square_V100_SXM2_final.csv};
\addplot table[x=dim,y=two_hybrid_opt,col sep=comma]   {plotdata/stra_square_V100_SXM2_final.csv};
\addplot table[x=dim,y=two_ref,col sep=comma]          {plotdata/stra_square_V100_SXM2_final.csv};

\end{axis}
\end{tikzpicture}
\begin{tikzpicture}[scale=1.05]
\begin{axis}[
    xlabel={$k = 4096$, $m = n$ vary},
    x label style={at={(axis description cs:0.5,0.02)}},
    ylabel={\emph{Effective} TFLOPS},
    y label style={at={(axis description cs:0.06,0.5)}},
    xmin=0,
    xmax=20000,
    ymin=0,
    ymax=18.5,
    xtick={1000,2000,3000,4000,5000,6000,7000,8000,9000,10000,12000,14000,16000,18000,20000},
    ytick={2,4,6,8,10,12,14,15.667,18.5},
    scaled x ticks=false,
    scaled x ticks=base 10:-3,
    grid=major,
    axis background/.style={fill=lightgray!20},
    mark size=1pt,
    cycle list name=jianyugpucolor,
    restrict y to domain=1:inf,
    legend style={
        at={(0.99,0.01)},
        anchor=south east,
        legend columns=2,
        font=\tiny,
        rounded corners=1pt,
        nodes={scale=1.1, transform shape},
        cells={anchor=west},
    },
    legend entries = {
        \cutlass{}\\
        \cublas{} \\
        $ 1~level, Ours $\\
        $ 1~level, Reference $\\
        $ 2~level, Hybrid $\\
        $ 2~level, Reference $\\
    },
    ]

\addplot table[x=dim,y=cutlass,col sep=comma]        {plotdata/stra_fixk4096_V100_SXM2.csv};
\addplot table[x=dim,y=cublas,col sep=comma]         {plotdata/stra_fixk4096_V100_SXM2.csv};
\addplot table[x=dim,y=one_abc_opt,col sep=comma]    {plotdata/stra_fixk4096_V100_SXM2.csv};
\addplot table[x=dim,y=one_ref,col sep=comma]        {plotdata/stra_fixk4096_V100_SXM2.csv};
\addplot table[x=dim,y=two_hybrid_opt,col sep=comma] {plotdata/stra_fixk4096_V100_SXM2.csv};
\addplot table[x=dim,y=two_ref,col sep=comma]        {plotdata/stra_fixk4096_V100_SXM2.csv};

\end{axis}
\end{tikzpicture}
\begin{tikzpicture}[scale=1.05]
\begin{axis}[
    xlabel={$m = n = 8192$, $ k $ vary},
    x label style={at={(axis description cs:0.5,0.02)}},
    ylabel={\emph{Effective} TFLOPS},
    y label style={at={(axis description cs:0.06,0.5)}},
    xmin=0,
    xmax=20000,
    ymin=0,
    ymax=18.5,
    xtick={1000,2000,3000,4000,5000,6000,7000,8000,9000,10000,12000,14000,16000,18000,20000},
    ytick={2,4,6,8,10,12,14,15.667,18.5},
    scaled x ticks=false,
    scaled x ticks=base 10:-3,
    grid=major,
    axis background/.style={fill=lightgray!20},
    mark size=1pt,
    cycle list name=jianyugpucolor,
    restrict y to domain=1:inf,
    legend style={
        at={(0.99,0.01)},
        anchor=south east,
        legend columns=2,
        font=\tiny,
        rounded corners=1pt,
        nodes={scale=1.1, transform shape},
        cells={anchor=west},
    },
    legend entries = {
        \cutlass{} \\
        \cublas{} \\
        $ 1~level, Ours $\\
        $ 1~level, Reference $\\
        $ 2~level, Hybrid $\\
        $ 2~level, Reference $\\
    },
    ]

\addplot table[x=dim,y=cutlass,col sep=comma]        {plotdata/stra_rankk8192_V100_SXM2.csv};
\addplot table[x=dim,y=cublas,col sep=comma]         {plotdata/stra_rankk8192_V100_SXM2.csv};
\addplot table[x=dim,y=one_abc_opt,col sep=comma]    {plotdata/stra_rankk8192_V100_SXM2.csv};
\addplot table[x=dim,y=one_ref,col sep=comma]        {plotdata/stra_rankk8192_V100_SXM2.csv};
\addplot table[x=dim,y=two_hybrid_opt,col sep=comma] {plotdata/stra_rankk8192_V100_SXM2.csv};
\addplot table[x=dim,y=two_ref,col sep=comma]        {plotdata/stra_rankk8192_V100_SXM2.csv};

\end{axis}
\end{tikzpicture}
\vspace{-0.2in}
\caption{Performance of various \strassen{} implementations on V100 with single precision: the \textbf{x-axis} denotes the matrix size, and the \textbf{y-axis} denotes the floating point efficiency in \texttt{TFLOPS}. Our 1-level and hybrid 2-level implementations are built on \cutlass{}, while the reference implementations are linked with \cublas{}.
}
\label{fig:stra_square}
\end{figure}


\noindent\textbf{Result:}
In \figref{fig:stra_square}, we report the single precision floating point efficiency of \cublas{}, \cutlass{}, and various \strassen{} implementations on a V100 GPU.
The 1-level and 2-level reference implementations~\cite{StrassenGPU2} are linked with cuBLAS 9.1.
For the 2-level hybrid implementation, we use reference implementation at the top level and our 1-level implementation at the bottom level.



By comparing the performance of various implementations, we make the following observations:
\begin{itemize}[leftmargin=*]
    \item For 1-level, our \strassen{} implementation outperforms \cutlass{} and \cublas{} when the problem sizes $ m=n=k $ are as small as 1,536. The reference implementation cannot get the comparable performance with our implementation until the problem sizes are larger than 10,000.
    For 2-level, our hybrid implementation outperforms the reference implementation. 
    \item Our implementation has the same memory consumption as \cutlass{}, while the 1-level reference implementation consumes much more memory.
    With V100 GPU (16 GB global memory), our 1-level \strassen{} can compute matrix multiplication for square problem sizes as large as 36,000, while the reference implementation runs out of memory after reaching 22,500.
    \item Our 1-level and hybrid 2-level \strassen{} implementations achieve the best performance over the entire spectrum of problem sizes compared to the reference implementations, with no or less additional memory consumption.
    Our hybrid 2-level implementation can get up to $1.22\times$ (ideally $1.3\times$) speedup compared to \cutlass{}  and $1.19\times$ speedup compared to \cublas{} when $ m = n = k = 20,480 $.
\end{itemize}
In summary, our 1-level \strassen{} algorithm can achieve practical speedup even 
for small (say $<3,000$) and non-square matrices without using any extra workspace. 
As a result, our methods can easily benefit different matrix shapes and be applied to different applications
such as matrix decomposition and tensor contraction.
For large problem sizes ($>9,000$), our hybrid 2-level \strassen{} algorithm 
can further provide speedup over our 1-level algorithm with additional $\mathcal{O}(mk/4+kn/4+mn/4)$ workspace.






\section{Analysis} \label{s:analysis} \begin{figure}[!t]
\centering
{
\setlength{\tabcolsep}{15pt}
\renewcommand{\arraystretch}{1.5}
  \begin{tabular}{ c | l | r }
  \whline
  Notation & Description & Value \\
  \whline
$ \tau_\texttt{flop} $ &  Arithmetic operation throughput & 15.67 \texttt{TFLOPS}   \\
\hline
$ \tau_{\texttt{gmop}} $  & Global memory bandwidth & 1.08 \texttt{TMOPS}\\ 
\hline
$ \tau_{\texttt{smop}} $  & Shared memory bandwidth & 15.30 \texttt{TMOPS}\\ 
\hline
  $ T $     & Total execution time (in seconds) & \\
\hline
$ t_xt_y $ & Number of threads per thread block & $(m_Sn_S)/(m_Rn_R)$ \\ %
\hline
  $N_\texttt{flop}^{\times}$ & Total \texttt{flops} for \gemm{} per thread block & $2 m_S n_S k$ \\
\hline
  $ N_\texttt{flop}^{+}(A) $ & Extra $+$ operations for operand $A$ & $(W_A-1)m_S k$  \\
\hline
  $ N_\texttt{flop}^{+}(B) $ & Extra $+$ operations for operand $B$ & $(W_B-1)n_S k$  \\
\hline
  $ N_\texttt{flop}^{+}(C) $ & Extra $+$ operations for operand $C$ & $(W_C-1)m_S n_S$  \\
\hline
$ N_\texttt{flop} $   & Total \texttt{flops} per thread block & \eref{eqn:n_flop} \\
\hline
$ T_\texttt{flop} $ & Time for arithmetic operations & \eref{eqn:t_flop} \\
\hline
$ N_\texttt{gmop}(X) $ & Global \texttt{mops} per block for operand $X$ & Equations \eqref{eqn:AB_gr_rs} \eqref{eqn:C_rg}  \\
\hline
$ N_\texttt{gmop} $ & Global memory operations per block & \eref{eqn:n_gmop} \\
\hline
$ T_\texttt{gmop} $ & Time for global memory operations & \eref{eqn:t_gmop} \\
\hline
$ N_\texttt{smop}(X) $ & Shared \texttt{mops} per block for operand $X$ & Equations \eqref{eqn:AB_gr_rs} \eqref{eqn:AB_sr}\\
\hline
$ N_\texttt{smop} $ & Shared operations per block & \eref{eqn:n_smop} \\
\hline
$ T_\texttt{smop} $ & Time for shared memory operations & \eref{eqn:t_smop} \\
  \whline
  \end{tabular}
}
\caption{Notation table for performance analysis.}
\label{tab:notation}
\end{figure}

In this section, we analyze our performance results by deriving a performance model for
\gemm{} and different variants (\secref{s:register}) from \strassen{}. 
Performance modeling helps us select the right blocking parameters, predict the performance, and understand the computation and memory footprint of \gemm{} and different \strassen{} implementations.

\subsection{Notation and Assumptions}



We summarize the notation in \figref{tab:notation} and assume the same 
three-level memory hierarchy as discussed in \secref{s:cuda_model}.
For a thread block,
the data movement through the memory hierarchy includes the following primitives:\newline
(i) loading the $ A $ and $ B $ \emph{Tile} for $ k/k_S $ times from global memory to shared memory, which is further decomposed into two steps: prefetching from \underline{g}lobal memory to \underline{r}egister files (line~\ref{l:A_gr}--\ref{l:extra_mop2} in~\algref{a:gpu_stra}) and storing back from \underline{r}egister files to \underline{s}hared memory (line~\ref{l:A_rs}--\ref{l:B_rs}):
\begin{equation}
\begin{split}
N_\texttt{gmop}(A_\texttt{gr}) &= N_\texttt{smop}(A_\texttt{rs}) =  m_S k_S(k/k_S), \\
N_\texttt{gmop}(B_\texttt{gr}) &= N_\texttt{smop}(B_\texttt{rs}) =  n_S k_S(k/k_S). \\
\end{split}
\label{eqn:AB_gr_rs}
\end{equation}
(ii) loading the $ A $ and $ B $ \emph{Fragment} from \underline{s}hared memory to \underline{r}egister files (line~\ref{l:A_sr}--\ref{l:B_sr}):
\begin{equation}
\begin{split}
N_\texttt{smop}(A_\texttt{sr}) &=  t_x t_y m_R k_S(k/k_S), \\
N_\texttt{smop}(B_\texttt{sr}) &=  t_x t_y n_R k_S(k/k_S). \\
\end{split}
\label{eqn:AB_sr}
\end{equation}
(iii) writing back the $ C $ \emph{Accumulator} from \underline{r}egister files to \underline{g}lobal memory (line~\ref{l:C_rg}):
\begin{equation}
N_\texttt{gmop}(C_\texttt{rg}) =  m_S n_S.
\label{eqn:C_rg}
\end{equation}
The total number of arithmetic operations for one thread block,
$ N_\texttt{flop} $, can be decomposed into matrix multiplications $ N_\texttt{flop}^\times $ and extra matrix additions 
\begin{equation} \label{eqn:n_flop}
     N_\texttt{flop} = N_\texttt{flop}^{\times} + N_\texttt{flop}^{+}(A) + N_\texttt{flop}^{+}(B) + N_\texttt{flop}^{+}(C).
\end{equation}
Due to the prefetching pipeline,
memory operations (handled by memory units)
are overlapped with the arithmetic operations (handled by CUDA cores).
We do not consider L1/L2 hardware cache effect, but we do take the read-only cache (texture memory) effect into account. 
We also do not consider the impacts of the task parallelism.

\subsection{Blocking Parameter Selection}
\label{s:blocking_choose}
Similar to \cite{GPUmodel1, GPUmodel2}, we select the blocking parameters
for \gemm{} and different \strassen{} variants (\secref{s:register})
by analyzing the hardware constraints such as
the maximum register number per thread
and the memory bandwidth.
Note that the following analysis mainly applies to large problem sizes when all SMs on V100 are fully utilized.
We assume $\tau_\texttt{flop} = 15.67$ \texttt{TFLOPS}~(\secref{s:volta}),  $ \tau_\texttt{gmop} = 1.08$ \texttt{TMOPS}\footnote{Due to the read-only cache (texture memory) effect, the global memory bandwidth is enhanced by a factor of 20\%, i.e., 900 (GB/s) $\times$ (1+20\%).},  $ \tau_\texttt{smop} = 15.30$ \texttt{TMOPS}\footnote{80 (\# SM) $\times$ 32 (\# banks/SM) $\times$ 4 (\# bank width: Bytes) $\times$ 1530 MHz~\cite{volta2}.},  $ m_S = n_S $, and $ m_R = n_R $ for square matrice cases.
The bounds for the blocking sizes are loose.

\noindent\textbf{Global memory bandwidth upper bound:}
Each thread block computes $ N_\texttt{flop} $ arithmetic operations and reads 
\begin{equation} \label{eqn:n_gmop}
    N_\texttt{gmop} = N_\texttt{gmop}(A_\texttt{gr})W_A + N_\texttt{gmop}(B_\texttt{gr})W_B +
    N_\texttt{gmop}(C_\texttt{gr})W_C
\end{equation} 
words. We can derive the bounds of $ m_S $ and $ n_S $ as
\begin{equation} \label{eqn:constrain_glb}
  (N_\texttt{flop}/N_\texttt{gmop}) \geq \texttt{sizeof(float)} (\tau_\texttt{flop}/\tau_\texttt{gmop}).
\end{equation}
It can be shown that $ m_S = n_S \geq 58.2 $, which results in the "Large" and "Huge" strategies for \gemm{}.
For 1-level \strassen{} where the total reads may double (e.g., Var\#0 and \#1), we need to choose the "Huge" strategy where $ m_S = n_S = 128 $. 
For 2-level \strassen{}, the required block sizes can be up to four times large.
As a result, no strategy is suitable.


\noindent\textbf{Shared memory bandwidth upper bound:}
Similarly, each thread block reads and writes
\begin{equation} \label{eqn:n_smop}
    N_\texttt{smop} = N_\texttt{smop}(A_\texttt{sr}) + N_\texttt{smop}(A_\texttt{rs}) + N_\texttt{smop}(B_\texttt{sr}) + N_\texttt{smop}(B_\texttt{rs}).
\end{equation}
We can derive the bounds of block sizes $ m_R $ and $ n_R $ as 
\begin{equation} \label{eqn:constrain_sha_bdw}
  (N_\texttt{flop}/N_\texttt{smop}) \geq \texttt{sizeof(float)} (\tau_\texttt{flop}/\tau_\texttt{smop}).
\end{equation}
As a result, we can get $ m_R = n_R \geq 4.1 $.

\noindent\textbf{Register number per thread constraint:}
In \algref{a:gpu_stra},
each thread requires 
$m_{R}\times n_{R}$ registers 
for the accumulator, $(W_Am_{R}+ W_Bn_{R})$ for fetching and prefetching 
operands $A$ and $B$, and $2(m_{R} + n_{R})$ for double buffering
operands between shared memory and register files.\footnote{At least $ W_A + W_B + 5 $ additional registers are needed:
$ W_A + W_B $ registers to track $ A $, $ B $ in the global memory during prefetching (line~\ref{l:A_gr}--\ref{l:extra_mop2});
1 register to store the loop end condition;
2 registers to track $ A $, $ B $ in the shared memory when prefetching (line~\ref{l:A_sr}--\ref{l:B_sr});
2 registers to track $ A $, $ B $ in the shared memory for storing back (line~\ref{l:A_rs}--\ref{l:B_rs}).
}
Since the maximum registers per thread is 255, $ m_R $ and $ n_R $ are bounded by 
\begin{equation} \label{eqn:constrain_reg}
    m_R n_R + (2 + W_A) m_R + (2 + W_B) n_R < 255.
\end{equation}
We can get $ m_R = n_R < 12 $.



\noindent\textbf{Shared memory size per SM constraint:}
Each thread block keeps the \{$ A $, $ B $\} \emph{Tile} in the shared memory, 
which requires
\begin{equation} \label{eqn:constrain_sha_size}
     \texttt{sizeof(float)} ( m_S k_S + n_S k_S )  < 96K,
\end{equation}
since the shared memory capacity per SM is 96 KB.

\noindent\textbf{Global memory prefetching precondition:}
Each thread prefetches one subcolumn of $ A $ with height $ m_R $ (line~\ref{l:A_gr}) and one subrow of $ B $ with width $ n_R $ (line~\ref{l:B_gr}), 
all $ t_x \times t_y $ threads in one thread block need to store back to the $ m_S k_S $ $ A $ \emph{Tile} (line~\ref{l:A_rs}) and the $ n_S k_S $ $ B $ \emph{Tile} (line~\ref{l:B_rs}),
so it requires
\begin{equation} \label{eqn:constrain_sha_pre}
    m_R t_x t_y \geq m_S k_S,
    n_R t_x t_y \geq n_S k_S. \\
\end{equation}
We can therefore get 
$ k_S \leq m_S / m_R $,
$ k_S \leq n_S / n_R $.

Basically, the Huge strategy in \cutlass{} (\secref{s:cutlass}) meets the bound requirement to maximize the performance for both \gemm{} and different variants from 1-level \strassen{} (Var\#0-\#3) on large problem sizes.

\subsection{Performance Prediction}
\label{s:perf_predict}
The total execution time $ T $ can be estimated as the maximum of the time of arithmetic operations $ T_\texttt{flop}$, the shared memory operations $ T_\texttt{smop}$, and the global memory operations $T_\texttt{gmop}$.
That is, $T=\max( T_\texttt{flop}, T_\texttt{gmop}, T_\texttt{smop} ) $.



\noindent\textbf{Arithmetic operations:}
We assume that the computation power of a GPU is split evenly among all active thread blocks,
i.e., each active thread block can get a portion of the peak throughput $\tau_\texttt{flop}$ of the whole GPU device: $\tau_\texttt{flop} / \#blocks $.
Here \#blocks is the maximum active thread blocks on one V100 device,
which is computed by
\begin{equation}
    \#blocks = \#SM \times \#max\_active\_blocks\_per\_SM\footnote{$ \#max\_active\_blocks\_per\_SM $ denotes
the number of the maximum active thread blocks per SM, which can be
returned from function \texttt{cudaOccupancyMaxActiveBlocksPerMultiprocessor}, or calculated with the CUDA Occupancy Calculator provided by NVIDIA~\cite{occupancy}. For Huge,
\gemm and all variants: 2;
For Small, \gemm{}: 24, 1-level Var\#0: 20, 2-level Var\#0: 18.
}.
\end{equation}
As a result, for $\lceil \frac{m}{m_S} \rceil \lceil \frac{n}{n_S} \rceil$ submatrix blocks, the total arithmetic operation time is
\begin{equation} \label{eqn:t_flop}
    T_\texttt{flop} = \left\lceil \frac{ \lceil \frac{m}{m_S} \rceil \lceil \frac{n}{n_S} \rceil  }{ \#blocks } \right\rceil \left(\frac{\#blocks\times N_\texttt{flop}}{ \tau_\texttt{flop} }\right)
\end{equation}
for $\#blocks$ active thread blocks.





\noindent\textbf{Shared memory operations:}
Similarly, we assume that the bandwidth of shared memory is allocated evenly
to each active thread block.
Given that the number of shared memory operations per thread block in~(\ref{eqn:n_smop}),
the total time spent on shared memory operations is
\begin{equation}
{
T_\texttt{smop} = \left\lceil \frac{ \lceil \frac{m}{m_S} \rceil \lceil \frac{n}{n_S} \rceil  }{ \#blocks } \right\rceil \left( \frac{ \texttt{sizeof(float)} \#blocks\times N_\texttt{smop} }{ \tau_{smop} }\right).
}
\label{eqn:t_smop}
\end{equation}
\noindent\textbf{Global memory operations:}
The global memory is accessible by all threads on all SMs and resides on the device level, so the bandwidth is not necessarily divided evenly by all thread blocks.\footnote{On the hardware layer, the HBM2 memory is connected to the chips through eight memory controllers in four memory stacks~\cite{volta3}, not coupled with individual SMs.}
Given the number of global memory operations per thread block in (\ref{eqn:n_gmop}),
the total time spent on global memory operations is
\begin{equation} \label{eqn:t_gmop}
    T_\texttt{gmop} = \lceil \frac{m}{m_S} \rceil \lceil \frac{n}{n_S} \rceil \left(\frac{ \texttt{sizeof(float)} N_\texttt{gmop}  }{ \tau_\texttt{gmop} }\right).
\end{equation}


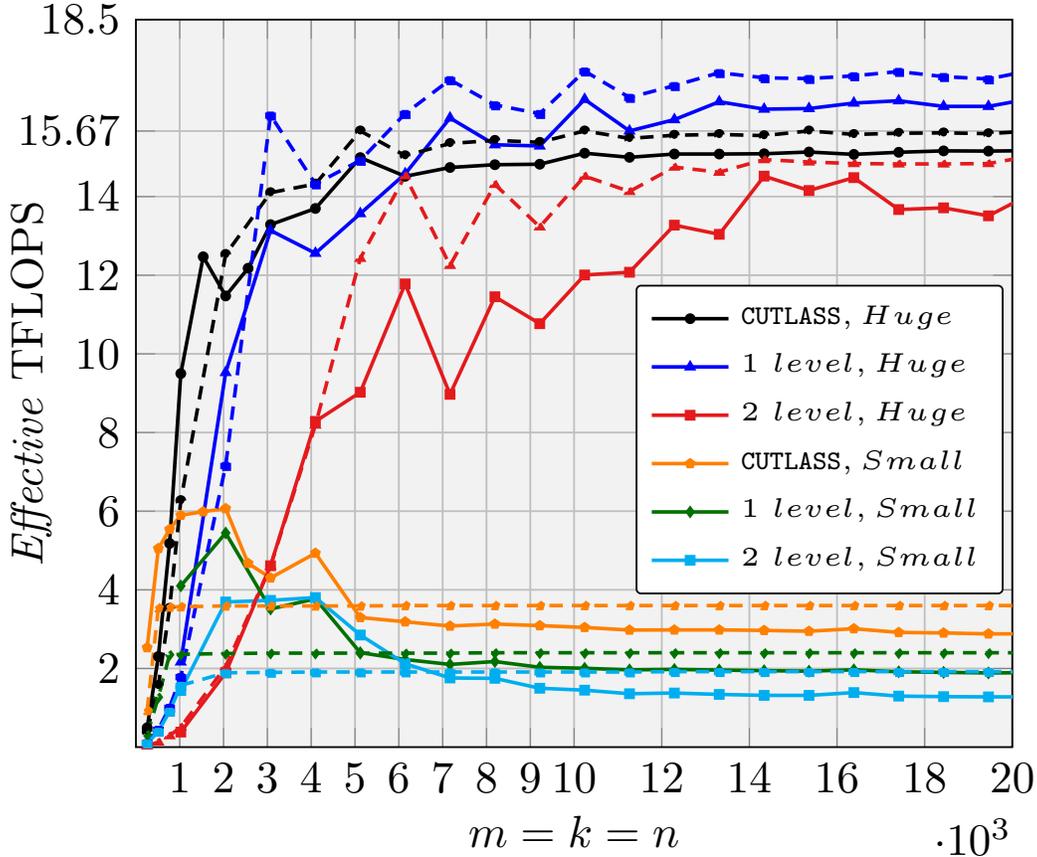
\begin{figure}[!t]
\center
\begin{tikzpicture}[scale=1.7]
\begin{axis}[
    xlabel={ $m = k = n$ },
    x label style={at={(axis description cs:0.5,0.02)}},
    ylabel={\emph{Effective} TFLOPS},
    y label style={at={(axis description cs:0.06,0.5)}},
    xmin=0,
    xmax=20000,
    ymin=0,
    ymax=18.5,
    xtick={1000,2000,3000,4000,5000,6000,7000,8000,9000,10000,12000,14000,16000,18000,20000},
    ytick={2,4,6,8,10,12,14,15.667,18.5},
    scaled x ticks=false,
    scaled x ticks=base 10:-3,
    grid=major,
    axis background/.style={fill=lightgray!20},
    mark size=0.8pt,
    cycle list name=jianyuhugecolor,
    restrict y to domain=0:inf,
    legend style={
        at={(0.99,0.21)},
        anchor=south east,
        legend columns=1,
        font=\tiny,
        rounded corners=1pt,
        nodes={scale=1.2, transform shape},
        cells={anchor=west},
    },
    legend entries = {
        $ \cutlass{}, Huge $\\
        $ 1~level, Huge $\\
        $ 2~level, Huge $\\
        $ \cutlass{}, Small  $\\
        $ 1~level, Small $\\
        $ 2~level, Small $\\
    },
    ]

\addplot table[x=dim,y=Huge,col sep=comma]   {plotdata/cutlass_perf_V100_SXM2_final.csv};
\addplot table[x=dim,y=Huge,col sep=comma]   {plotdata/stra_perf_square.csv};
\addplot table[x=dim,y=Huge,col sep=comma]   {plotdata/stra_perf_square2.csv};
\addplot table[x=dim,y=Small,col sep=comma]   {plotdata/cutlass_perf_V100_SXM2_final.csv};
\addplot table[x=dim,y=Small,col sep=comma]   {plotdata/stra_perf_square.csv};
\addplot table[x=dim,y=Small,col sep=comma]   {plotdata/stra_perf_square2.csv};

\addplot table[x=dim,y=gemm_huge,col sep=comma]   {plotdata/model_square.csv};
\addplot table[x=dim,y=stra_huge,col sep=comma]   {plotdata/model_square.csv};
\addplot table[x=dim,y=stra_huge,col sep=comma]   {plotdata/model_square2.csv};
\addplot table[x=dim,y=gemm_small,col sep=comma]   {plotdata/model_square.csv};
\addplot table[x=dim,y=stra_small,col sep=comma]   {plotdata/model_square.csv};
\addplot table[x=dim,y=stra_small,col sep=comma]   {plotdata/model_square2.csv};

\end{axis}
\end{tikzpicture}
\caption{
    Actual (solid line) and modeled (dashed line) performance of \cutlass{} and \strassen{} with Small and Huge strategies of block sizes.
}
\label{fig:model_square}
\end{figure}

We can predict the run time performance of various implementations, based on this performance model.
In \figref{fig:model_square}, we present the modeled and actual performance of \gemm{} and direct 1/2-level \strassen{} (\secsref{s:gpu_stra} and \ref{s:2level})
for square matrices with Huge and Small strategies of block sizes (\tabref{tab:gpu_block_size}).
The direct 1- and 2-level \strassen{} are implemented using 7/49 instances of different variants sequentially, without inter-kernel task parallelism (\secref{s:task}).

\subsection{Discussion and Analysis}
\label{s:perf_discuss}


\noindent\textbf{Impacts of the variants in \strassen{}:}
From our model, the performance differences between the variants~(\secref{s:register})
are determined by the operand counts $ W_{\{A, B, C\}} $, which mainly affects the number of global memory operations~(\ref{eqn:constrain_glb}) and $ T_\texttt{gmop} $, the total number of arithmetic operations $ N_\texttt{flop} $, and the register number~(\ref{eqn:constrain_reg}).
For example, comparing Var\#0 in 1-level \strassen{} with \gemm{}, we can find that the global memory operation number doubles, and the required register number increases by $ m_R + n_R $.

\noindent\textbf{Limitations and possible solutions:}
Our 2-level \strassen{} primitives may increase operands count
$W_A$, $W_B$, and $W_C$ up to four times.
These primitives may require up to
160 registers per thread by \eqref{eqn:constrain_reg}, and up to 1,900 GB/s global and
texture memory throughput by \eqref{eqn:constrain_glb}.
Regarding the current architecture, memory operations cannot be fully overlapped with the computations and registers must be
spilled to maintain two active thread block per SM (or just maintain one active thread block).
These two limitation factors suggest possible hardware 
improvements on future generation GPUs to make the 
2-level primitives practical.

Extra registers $next1_A$ and $next1_B$ in~\algref{a:gpu_stra} 
are used to prefetch extra operands at line~\ref{l:A_gr}--\ref{l:extra_mop2}, which are handled solely by the memory units thus overlapped
with rank-$k_S$ update during line~\ref{l:A_sr}--\ref{l:accum}.
For 2-level \strassen{}, the extra registers required for prefetching will exceed the constraint.
Moving arithmetic operations
at line~\ref{l:A_rs}--\ref{l:B_rs} to line~\ref{l:A_gr}--\ref{l:extra_mop2}
can reduce the register requirement
by reusing $next1_A$ and $next1_B$
but result in CUDA cores waiting for the memory
access,
thus decrease the number of overlapped memory operations.
Given that the 2-level \strassen{} primitives already require 
much higher memory bandwidth, it is not practical to trade overlapped memory operations with more registers.

To alleviate the register pressure and memory traffic, our 
\strassen{} primitives are good examples that could benefit
from Processing-In-Memory (PIM)~\cite{bennett2012combining,ahn2015pim,loh2013processing}.
With extended memory instructions that directly compute the 
arithmetic operations at line~\ref{l:A_rs}--\ref{l:B_rs} during the
fetching process at line~\ref{l:A_gr}--\ref{l:extra_mop2}, 
it is possible to remove all extra registers for prefetching.
The computation is done in-transit of the loading process, which
may also relieve the memory traffic in the memory hierarchy and reduce the required memory throughput.


\noindent\textbf{Cache effects:}
For the Small strategy, the actual performance is better than the modeled performance during the ramp-up stage. 
This shows the L1/L2 cache effects as there are two performance ``falling edges'' for the actual performance,
which are not captured by our performance model.




\section{Conclusion}
\label{s:gpu_conclusion}



We have presented a practical implementation of Strassen's algorithm on GPUs, which outperforms the state-of-the-art implementation on small problem sizes and consumes no additional memory compared to \gemm{}.
By developing a specialized kernel, we utilized the memory and thread hierarchies on GPUs.
By reusing the shared memory to store the temporary matrix sum during the packing process and the register files to hold the temporary matrix product during the accumulation process, we avoided the extra workspace requirement and reduced the additional memory movement. 
Besides the intra-kernel parallelism across the thread blocks, warps, and threads similar to \gemm{} implementation on GPUs, we also exploited the inter-kernel parallelism and batched parallelism and overlapped the bandwidth limited operations with the computation bound operations.
We demonstrated performance benefits for small and non-square matrices on a most recent Volta GPU, and verified the performance results by building an accurate performance model to choose the appropriate block sizes and predict the run time performance.
Together, we achieved both less memory and more parallelism with our customized kernels.
In the future, we will extend this work to other applications on GPUs, such as fast matrix multiplication algorithms~\cite{Benson15,FMM:IPDPS17}, high-dimensional tensor contractions~\cite{TC:Devin}, and convolution neural network~\cite{conv1,conv2}.





%

\section*{Acknowledgments}

This work was sponsored in part by the National Science Foundation under grant number CCF-1714091 and the 2016 Peter O'Donnell Distringuished Research Award.
This work was also made possible, in part, through HPC time donated by Microway, Inc. We gratefully acknowledge Microway for providing access to their GPU-accelerated compute cluster.
Access to the Maverick2 supercomputers administered by TACC is gratefully acknowledged.
We thank 
the rest of the SHPC team ({\tt http://shpc.ices.utexas.edu}) for their supports.

{\em Any opinions, findings, and conclusions or recommendations expressed in this material are those of the author(s) and do not necessarily reflect the views of the National Science Foundation. }

\bibliographystyle{plain}

\bibliography{biblio}


\end{document}